\documentclass[aps,prd,twocolumn,superscriptaddress]{revtex4-2}

\usepackage[dvips]{graphicx}
\usepackage{ulem}
\usepackage[usenames]{color}
\usepackage{amsmath}
\usepackage{amssymb}
\usepackage{bm}
\usepackage{mathrsfs}
\usepackage{amsbsy}
\usepackage{hyperref}
\usepackage[all]{hypcap}
\usepackage{longtable}
\usepackage{url}
\usepackage{multirow}

\begin{document}

\title{Mass spectrum of fully charmed $[cc][\bar c\bar c]$ tetraquarks}

\author{You-You Lin}

\author{Ji-Ying Wang}

\author{Ailin Zhang}
\email{zhangal@shu.edu.cn}
\affiliation{Department of Physics, Shanghai University, Shanghai 200444, China}

\begin{abstract}
There are three $S$ and seven $P$ fully charmed $[cc][\bar c\bar c]$ tetraquarks, the mass spectrum from $1S$ to $2P$ excitations is calculated in a non-relativistic quark potential model. In the calculation, the interactions among four internal quarks/antiquark are approximated as a dominant color interaction between a diquark and an antidiquark, and a residual interaction responsible for the diquark/antidiquark cluster effect. The color interaction between the diquark and the antidiquark is characterized by the conventional Cornell potential, while the residual interaction is modeled as the Yukawa-type scalar $\sigma$ and vector $\omega$ boson exchange potentials. In the numerical results, though the scalar $\sigma$ and vector $\omega$ boson exchange interactions reduce the masses of $S-$wave $[cc][\bar c\bar c]$ tetraquarks $40-50$ MeV, they contribute to the masses of other $[cc][\bar c\bar c]$ tetraquarks small. The mass splittings of $[cc][\bar c\bar c]$ tetraquarks between different multiplets and within the same multiplet are smaller than those in charmonium, and the splittings are affected by the scalar $\sigma$ or vector $\omega$ boson exchange interactions small. Our calculations suggest that the observed $X(6600)$, $X(6900)$ and $X(7300)$ should be different radial excitations of $[cc][\bar c\bar c]$. The measurements of the $J^{PC}$ quantum numbers and the mass splittings will be helpful to identify and understand of the $[cc][\bar c\bar c]$ candidates.
\end{abstract}

\maketitle

\section{Introduction}\label{intro}
Four quark states have been proposed for a long time~\cite{GellMann1964}, and have been extensively explored since the observation of charmed $X(3872)$ by Belle Collaboration in 2003~\cite{Belle:2003nnu}. Though QCD accounts for the interactions in multiquark states, the properties of multiquark states can not be derived directly from QCD for the non-perturbational interaction and confinement. It is also difficult to learn the explicit internal quark structure and dynamics in these many body systems. The internal quark structure and dynamics in multiquark states are still not clear and require more exploration.

There may be different quark correlations or clusters, and hence configurations, inside multi-quark states, it is the same case in four quark states. For example, four-quark states with charm quarks may have tetraquark, molecule, hadron-charmonium~\cite{Dubynskiy:2008mq} or adjoint charmonium configurations~\cite{Braaten:2013boa}. A $[qq][\bar q\bar q]$ tetraquark state consists of two quarks and two anti-quarks, where the two quarks make a $[qq]$ diquark and the two anti-quarks make a $[\bar q\bar q]$ antidiquark. Fully charmed $[cc][\bar c\bar c]$ tetraquarks with four charm quarks/anti-quarks provide an ideal platform to study the dynamics of multi-quark states in a non-relativistic limit.

In $2020$, fully charmed resonances with a narrow structure labeled $X(6900)$ and a broad structure in the mass region $6.2-6.8$ GeV were first announced by LHCb~\cite{LHCb:2020bwg}. Subsequent experiments of CMS~\cite{CMS:2023owd} and ATLAS~\cite{ATLAS:2023bft} collaborations not only confirmed the existence of $X(6900)$, but also reported two new resonances $X(6600)$ and $X(7300)$.

The mass of $X(6900)$ is measured by LHCb~\cite{LHCb:2020bwg}
\begin{equation}
M_1=6905\pm11\pm7~\mathrm{MeV}
\end{equation}
without interference with a non-resonant single-parton scattering (NRSPS) continuum. With the interference incorporated, the mass is shifted to
\begin{equation}
M_1=6886\pm11\pm11~\mathrm{MeV}.
\end{equation}
The masses are measured by CMS with Breit-Wigner lineshape~\cite{CMS:2023owd}
\begin{equation}
\begin{aligned}
&X(6600):~M=6552 \pm 10 \pm12~\mathrm{MeV},\\
&X(6900):~M=6927 \pm 9 \pm4~\mathrm{MeV},\\
&X(7300):~M=7287^{+20}_{-18} \pm5~\mathrm{MeV}.	
\end{aligned}
\end{equation}
With possible interferences incorporated, the masses are fitted with
\begin{equation}
\begin{aligned}
&X(6600):~M=6638^{+43+16}_{-38-31}~\mathrm{MeV},\\
&X(6900):~M=6847^{+44+48}_{-28-20}~\mathrm{MeV},\\
&X(7300):~M=7134^{+48+41}_{-25-15}~\mathrm{MeV}.	
\end{aligned}
\end{equation}

Two fit models A (three resonances with interferences) and B (two resonances with the lower broad
structure interfering with the SPS background) similar to those in the analysis of LHCb~\cite{LHCb:2020bwg} are also performed by ATLAS~\cite{ATLAS:2023bft}. Accordingly, three resonances were reported in model A in di-$J/\psi$ channel with
\begin{equation}
\begin{aligned}
&M_0=6.41 \pm 0.08^{+0.08}_{-0.03}~\mathrm{GeV},\\
&M_1=6.63\pm0.05^{+0.08}_{-0.01}~\mathrm{GeV},\\
&M_2=6.86\pm0.03^{+0.01}_{-0.02}~\mathrm{GeV},\\
\end{aligned}
\end{equation}
and one resonance in $J/\psi \psi(2S)$ channel with
\begin{equation}
M_3=7.22\pm0.03^{+0.01}_{-0.04}~\mathrm{GeV}.
\end{equation}
In model B, two resonances were reported in di-$J/\psi$ channel with
\begin{equation}
\begin{aligned}
&M_0=6.65 \pm 0.02^{+0.03}_{-0.02}~\mathrm{GeV},\\
&M_2=6.91\pm0.01\pm0.01~\mathrm{GeV},\\
\end{aligned}
\end{equation}
and one resonance in $J/\psi \psi(2S)$ channel with
\begin{equation}
M_3=6.96\pm0.05\pm0.03~\mathrm{GeV}.
\end{equation}
In the analysis of CMS, the measured mass has an uncertainty up to $150$ MeV in different fit models~\cite{CMS:2023owd}. In the analysis of ATLAS~\cite{ATLAS:2023bft}, the measured mass has an uncertainty up to $260$ MeV in different fit models. So far, the measurements have large uncertainties.

Theoretical study of fully charmed $[cc][\bar c\bar c]$ tetraquark has started over $40$ years~\cite{Iwasaki:1975pv,Chao:1980dv,PhysRevD.25.2370,Zouzou1986}. Since the discovery of $X(6900)$, its possibility as a fully charmed $[cc][\bar c\bar c]$ tetraquark has been studied with different methods, such as QCD sum rule, Bethe-Salpeter equation, potential model, coupled channel analysis, string junction model, etc~\cite{Zhang:2020xtb,Wang:2020ols,Li:2021ygk,Ke:2021iyh,Zhu:2020xni,Bedolla:2019zwg,Liu:2021rtn,Dong:2020nwy,Wang:2022yes}. In these studies, it is believed that fully charmed quarks/anti-quarks can hardly make a molecular state since there exists no proper potential to bind the charmonium pairs, and the $X(6900)$ is supposed a fully charmed tetraquark.

In constituent quark model, the mass spectrum of fully charmed $[cc][\bar c\bar c]$ tetraquarks has been computed in the diquark-antidiquark picture or the compact four bodies picture~\cite{Karliner:2016zzc,Wu:2016vtq,Anwar:2017toa,Debastiani:2017msn,Liu:2019zuc,Lu:2020cns,Jin:2020jfc,Giron:2020wpx,Liu:2021rtn,Zhang:2022qtp,Wang:2023jqs}. In a non-relativistic diquark-antidiquark model with Coulomb-plus-linear potential between a diquark and an antidiquark~\cite{Debastiani:2017msn}, the mass of $[cc][\bar c\bar c]$ is predicted with $5969.4-6997.7$ MeV for the $1S$ to the $2P$ state. In a dynamical diquark-antidiquark picture~\cite{Giron:2020wpx}, the non-orbital coupling is accounted only inside the diquark and the orbital coupling is considered between the diquark and the antidiquark. The masses of fully charmed $[cc][\bar c\bar c]$ tetraquarks are predicted around $6264.0-7013.0$ MeV from the $1S$ state to the $2P$ state, and $X(6900)$ was interpreted as a $2S$ state. In a relativistic diquark-antidiquark model with quasipotential approach, the internal structure of diquark is taken into account by including a diquark-gluon form factor. The masses of the $[cc][\bar c\bar c]$ are predicted with $6190$ MeV for the $0^{++}$ ground state, $6271$ MeV for the $1^{+-}$ ground state and $6367$ MeV for the $2^{++}$ ground state~\cite{Faustov:2020qfm}. The mass splitting in the ground state is about $81-177$ MeV. In a relativistic diquark-antidiquark model with Godfrey-Isgur (GI) potential, the predicted masses vary in the region $5883-7380$ MeV from the $1S$ to the $3D$ states~\cite{Bedolla:2019zwg}. In a compact four bodies picture, the internal interaction is simulated by an additive two bodies potential, and the mass spectrum is obtained higher, where the masses of $1S$ wave states are about $6.5$ GeV~\cite{Liu:2019zuc,Liu:2021rtn,Zhang:2022qtp}. In the compact four bodies picture, more states are predicted in comparison to the experiments and the mass splittings between states are relatively small.

In short, in those literatures, the predicted masses of $[cc][\bar c\bar c]$ tetraquarks from $1S$ state to $2P$ state have large difference, the relevant mass splittings are largely different either. Accordingly, the assignments of the experimental observations such as $X(6600)$, $X(6900)$ and $X(7300)$ are different. More prediction of the mass spectrum for the fully charmed $[cc][\bar c\bar c]$ tetraquarks is required.

Diquark was first mentioned in the original paper on quarks by Gell-Mann~\cite{GellMann1964}, and was later constructed to describe a baryon as a two bodies composite, a quark and a diquark, by Ida \textit{et al.}~\cite{Ida:1966ev} and Lichtenberg \textit{et al.}~\cite{Lichtenberg:1967zz}. Diquark was subsequently employed to construct tetraquarks, pentaquarks and other multiquark states~\cite{Jaffe:2003sg,Karliner:2003dt,Maiani:2004uc,Ferretti:2011zz,Maiani:2015vwa,Maiani:2017kyi}, more explorations could be found in some reviews~\cite{Anselmino:1992vg,Jaffe:2004ph,Chen:2016qju,Lebed:2016hpi,Olsen:2017bmm,Barabanov:2020jvn} and references therein.

As well known, two quarks in a hadron may attract each other and make up a diquark when the two quarks are in an antisymmetric color representation.
Once the diquark/antidiquark is treated as a point stuff, the tetraquark has a similar color structure to the normal $q\bar q$ meson. Therefore, the dominant strong interaction in tetraquarks is usually thought as the gluon-exchange interaction between the diquark and the antidiquark~\cite{Maiani:2004uc,Maiani:2004vq,Brodsky:2014xia}.

In many computations of the mass spectrum of tetraquarks in constituent quark potential model~\cite{Debastiani:2017msn,Bedolla:2019zwg,Tiwari:2021tmz}, the motion and interaction of charm quarks inside a diquark are incorporated in the constituent mass of diquark, and the effect of diquark size or other intricate dynamics has not been explicitly accounted in the effective potential between the diquark and the antidiquark. In fact, the diquark/antidiquark cluster in hadrons is a kind of strong correlation between pairs of quarks though its properties and evidence are still not clear. The diquark/antidiquark strong correlation effect, the cluster effect, can not be simply omitted. The interaction from the cluster effect plays a minor role in tetraquarks and can be regarded as a residual color interaction, which should be explicitly incorporated in the interaction potential.

Since Yukawa's assumption that the forces between nucleons
are mediated by the exchange of pion~\cite{Yukawa:1935xg}, many
boson-exchange potential models have been constructed
to describe the nucleon-nucleon/antinucleon interactions and relevant experimental behavior~\cite{Bryan:1964zzb,Ball:1965sa}. The boson-exchange models have led to a nice understanding of the interactions. Subsequently, field theories are successfully constructed to describe the interactions between $NN$ or $N \bar N$~\cite{Brown:1970mj,Partovi:1969wd,Machleidt:1987hj,Meissner:1987ge,Holzenkamp:1989tq,Weinberg:1991um}. In these investigations, hadrons were used as the elementary fields, and the theories are decisively model dependent. With the advent of quantum chromodynamics, effective quark theories with chiral symmetry and spontaneous breaking of chiral symmetry have been constructed~\cite{Manohar:1983md,Tornqvist:1993ng,Peter:1997me}. The nucleon-nucleon and nucleon-meson interactions can be ``derived'' from these effective theories and hence from the symmetry properties of quantum chromodynamics.

Two hadrons may make a hadronic molecule such as deuterons and four quark molecules. Four quark molecules consist of two quarks and two anti-quarks with a $(q\bar q)(q\bar q)$ configuration, where each $q\bar q$ cluster is in color singlet. The interaction between a pair of charmed mesons was first studied by Voloshin and Okun~\cite{Voloshin:1976ap}. The possibilities of the molecular states involving charmed quarks was also proposed. T\"ornqvist~\cite{Tornqvist:1991ks} studied the possible deuteronlike meson-meson states bound by pions. They found that the one-pion-exchange potential is likely to bind a few states composed of two ground state mesons. But the attraction between the mesons was found to be not strong enough to bind two light mesons such as $K\bar K^*$ alone. One-pion-exchange potential was then extended to the heavy meson sector~\cite{Manohar:1983md,Tornqvist:1993ng,Valcarce:2005em,Ding:2009vj,Lee:2009hy,Sun:2011uh,Wang:2013kva,Guo:2017jvc}, and many deuteronlike charmed states were found to be bound or nearly bound.

From the original hypothesis of Yukawa to modern effective field theories, the boson-exchange interactions work well in many physics process though this interaction cannot be derived directly from QCD. By intuition, the boson-exchange interaction indicate the right interaction between two color singlet clusters (nucleon-nucleon, nucleon-antinucleon, baryon-meson and meson-meson). Once the dominant color interaction has been taken into account, the boson-exchange interaction is employed phenomenlogically as an attempt to mock the cluster effect (residual interaction) in tetraquark though this boson-exchange interaction has been only employed to describe the interactions in molecules before.

Of course, we think it is a challenge to isolate the dominant color interaction and the residual interaction based on QCD or some effective field theories. In this paper, the boson-exchange interaction is incorporated in tetraquarks. The dominant gluon-exchange interaction is responsible for the main color interaction between the two colored $[cc]$ and $[\bar c\bar c]$, which is similar to that in normal $q\bar q$ mesons. The boson-exchange interaction is responsible for residual color interactions between the two colored cluster: $[cc]$ and $[\bar c\bar c]$. In this way, the mass spectrum of fully charmed $[cc][\bar c\bar c]$ tetraquarks was systematically computed.

The rest of this paper is organized as follows, the construction of the potential and wave function is presented in Sec.~\ref{model}. The mass spectrum was computed and presented in Sec.~\ref{results} with pertinent discussions. The last section is reserved for a summary.

\section{Framework}\label{model}
\subsection{Potential model}
In a non-relativistic quark potential model, the Hamiltonian of a fully charmed $[cc][\bar c\bar c]$ tetraquark could be written as that in Refs.~\cite{DeRujula:1975qlm,Eichten:1978tg}
\begin{equation}\label{eq:potential}
H=m_{d}+m_{\bar{d}}+\frac{p^2}{2\mu}+V,
\end{equation}
where $m_{d}$ and $m_{\bar{d}}$ are the constituent masses of diquark and antidiquark, respectively. $\frac{p^2}{2\mu}$ is the kinetic energy in c.m.s, determined by the relative momentum $p$ and reduced mass $\mu=\frac{m_dm_{\bar{d}}}{m_d+m_{\bar{d}}}$. $V$ is an effective quark potential within the tetraquark.

As analysed in the introduction, the dominant color interaction between the diquark and the antidiquark is similar to that in normal $q\bar q$ meson, and this interaction can be described by an effective quark potential. The residual color interaction from the cluster effect can be described by an effective one boson exchange (OBE) potential. Therefore the whole effective potential
\begin{equation}
V=V_{Cornell}+V_{OBE}.
\end{equation}
In our scheme, an explicit Cornell potential including both one gluon exchange interaction and a confinement interaction is employed~\cite{Eichten:1978tg}
\begin{align}\label{eq:cor}
V_{Cornell}=&V_{OGE}+V_{CON}\notag\\
=&k_s\left\{\frac{\alpha_s}{r}-\frac{\alpha_s}{m_dm_{\bar{d}}} \left[\frac{8\pi}{3} \boldsymbol{S}_{d} \cdot \boldsymbol{S}_{\bar{d} } {\delta}^3(\boldsymbol{r})\right.\right.\notag\\
&\left.+\frac{1}{r^3}\boldsymbol{S}_T\right]+\left(-\frac{\alpha_s}{2r^3}+\frac{3b}{8r}\right)\frac{\boldsymbol{L} \cdot \boldsymbol{S}_d}{m_d^2}\notag\\
&-\frac{\alpha_s}{r^3} \frac{\boldsymbol{L} \cdot \boldsymbol{S}_d}{m_d m_{\bar{d}}}+\left(-\frac{\alpha_s}{2r^3}+\frac{3b}{8r}\right)\frac{\boldsymbol{L} \cdot \boldsymbol{S}_{\bar{d}}}{m_{\bar{d}}^2}\notag\\
&\left.-\frac{\alpha_s}{r^3} \frac{\boldsymbol{L} \cdot \boldsymbol{S}_{\bar{d} }}{m_d m_{\bar{d}}}-\frac{3}{4}br\right\},
\end{align}
where $k_s$ is a color factor, depending on the color configuration of tetraquark states. $\boldsymbol{S}_T=\frac{3\boldsymbol{S}_d \cdot \boldsymbol{r} \boldsymbol{S}_{\bar{d} } \cdot \boldsymbol{r}}{r^2}-\boldsymbol{S}_d \cdot \boldsymbol{S}_{\bar{d}}$. To eliminate the singularity induced by the delta function in the equation, we replace it with a Gaussian function as did in Ref.~\cite{Godfrey:1985xj},
\begin{equation}
{\delta}^3(\boldsymbol{r})\rightarrow{\left(\frac{\sigma}{\sqrt{\pi}}\right)}^3 e^{-\sigma^2 r^2}.
\end{equation}
As the charmed diquark and antidiquark has the same mass, Eq.~(\ref{eq:cor}) is therefore simplified to
\begin{align}
V_{Cornell}=&k_s \left\{\frac{\alpha_s}{r}-\frac{\alpha_s}{m_d^2}\left[\frac{8\pi}{3}\boldsymbol{S}_d \cdot \boldsymbol{S}_{\bar d}{\left(\frac{\sigma}{\sqrt{\pi}}\right)}^3 e^{-{\sigma}^2 r^2}\right.\right.\notag\\
&\left.\left.+\frac{1}{r^3} \boldsymbol{S}_T\right]+\left(-\frac{3\alpha_s}{2r^3}+\frac{3b}{8r}\right)\frac{\boldsymbol{L} \cdot \boldsymbol{S}}{m_d^2}-\frac{3}{4}br\right\}.
\end{align}

For one boson exchange potential between a diquark and an antidiquark, the proper mediator should respect the conservation of isospin and angular momentum as those in hadronic molecular system. As the charmed diquark and antidiquark has isospin $0$, the isoscalar scalar $\sigma$ and vector $\omega$ are suitable to be the interaction mediator between the diquark and the antidiquark.

In our calculations, the boson exchange potentials between the charmed diquark and antidiquark are taken from Ref.~\cite{Liu:2019stu}. The $\sigma$ exchange potential is universal for all kinds of hadrons, while the $\omega$ exchange potential differs in different hadron systems.
\begin{equation}
V_\sigma=-g_\sigma^2m_\sigma W_Y(m_\sigma r,\frac{\Lambda}{m_\sigma}),
\end{equation}
\begin{align}
V_{\omega}=&-g_{\omega}^{2}m_{\omega}W_Y(m_\omega r,\frac{\Lambda}{m_\omega})-\frac{f_{\omega}^{2}}{4m_d^{2}}\left(-\frac{2}{3}\boldsymbol{S}_{d}\cdot\boldsymbol{S}_{\bar d}\right. \notag \\
&\times m_{\omega}^3 d(x,\lambda)+\frac{2}{3}\boldsymbol{S}_{d}\cdot\boldsymbol{S}_{\bar d}m_{\omega}^{3}W_Y(m_\omega r,\frac{\Lambda}{m_\omega})\notag\\
&\left.-\frac{1}{3}\boldsymbol{S}_T m_{\omega}^{3}W_T(m_\omega r,\frac{\Lambda}{m_\omega})\right),
\end{align}
where the form factors accounting for the size of $\sigma$ and $\omega$ bosons,
\begin{equation*}
d(x,\lambda)=\frac{(\lambda^2-1)^2}{2\lambda}\frac{e^{-\lambda x}}{4\pi},
\end{equation*}
\begin{equation*}
W_Y(x,\lambda)=W_Y(x)-\lambda W_Y(\lambda x)-\frac{(\lambda^2-1)}{2\lambda}\frac{e^{-\lambda x}}{4\pi},
\end{equation*}
\begin{equation*}
W_T(x,\lambda)=W_T(x)-\lambda^3W_T(\lambda x)-\frac{(\lambda^2-1)}{2\lambda}\lambda^2\left(1+\frac{1}{\lambda x}\right)\frac{e^{-\lambda x}}{4\pi},
\end{equation*}
with
\begin{equation*}
W_{Y}(x)=\frac{e^{-x}}{4\pi x},
\end{equation*}
\begin{equation*}
W_{T}(x)=\left(1+\frac{3}{x}+\frac{3}{x^{2}}\right)\frac{e^{-x}}{4\pi x}.
\end{equation*}
Similar boson exchange potentials have been employed to study $X(3872)$ in hadronic molecular picture in Refs.~\cite{Liu:2009qhy,Lee:2009hy}.

\subsection{Wave Function}
Internal structure and $J^{PC}$ quantum numbers of $[cc][\bar c\bar c]$ tetraquarks have been studied in many Refs.~\cite{Chao:1980dv,Wu:2016vtq,Debastiani:2017msn,Bedolla:2019zwg,Giron:2020wpx}. The wave function of $[cc][\bar c\bar c]$ is constructed as~\cite{Wu:2016vtq}
\begin{equation}\label{eq:psi}
	\Psi_{JM}=\chi_f\otimes\chi_c\otimes[\Psi_{lm}(\boldsymbol{r})\otimes \chi_s]_{JM},
\end{equation}
where $\chi_f$, $\chi_c$ and $\chi_s$ are the flavor, color and spin wave functions, respectively.

Diquark is supposed to have no internal spatial structure, and is antisymmetric in the space of the flavor, color and spin. For fully charmed tetraquarks,
\begin{equation}
\chi_{f}=\{cc\}\{\bar{c}\bar{c}\}
\end{equation}
is symmetric. Therefore, the product of the color and the spin wave function of each diquark, $\chi_c \chi_s$, should be antisymmetric. The diquark $[cc]$ may be in two available color representations, $\mathbf{\bar 3}_c$ and $\mathbf{6}_c$. Accordingly, the spin wave function must be symmetric and antisymmetric, respectively. Four wave functions are given as follows~\cite{Wu:2016vtq}
\begin{eqnarray}\label{eq:psics}
    & \chi_1=|\{cc\}^{\bar{3}}_{1}\{\bar{c}\bar{c}\}^{3}_{1}\rangle_0,\\
	&\chi_2=|\{cc\}^{\bar{3}}_{1}\{\bar{c}\bar{c}\}^{3}_{1}\rangle_1,\\
	&\chi_3=|\{cc\}^{\bar{3}}_{1}\{\bar{c}\bar{c}\}^{3}_{1}\rangle_2,\\
	&\chi_4=|\{cc\}^{6}_{0}\{\bar{c}\bar{c}\}^{\bar{6}}_{0}\rangle_0,
\end{eqnarray}
where a notation $|\{cc\}^{color}_{spin}\{\bar{c}\bar{c}\}^{color}_{spin}\rangle_{spin}$ is used. The first three wave functions are in $\mathbf{\bar 3}_c\otimes \mathbf{3}_c$ color configuration, while the last is in $\mathbf{6}_c\otimes \mathbf{\bar 6}_c$ color configuration.

In Ref.~\cite{Debastiani:2017msn}, the sextet diquark has been found disfavored to make a tetraquark in the diquark-antidiquark picture. In a four bodies picture, it is found that the sextet diquark usually mix with an antitriplet diquark, and cannot form a tetraquark state independently~\cite{Park:2013fda,Wu:2016vtq,Liu:2019zuc}. In fact, the one gluon exchange potential for sextet diquark is repulsive with a color factor $+1/3$. Therefore, only the anti-triplet diquark $[cc]$ and the triplet antidiquark $[\bar c\bar c]$ are suitable to construct the tetraquarks.

The spatial wave function $\Psi_{lm}(\boldsymbol{r})$ is solved using Gaussian expansion method (GEM)~\cite{Hiyama:2003cu}. In GEM, $\Psi_{lm}(\boldsymbol{r})$ is expanded with a series of Gaussian functions,
\begin{align}
\Psi_{lm}(\boldsymbol{r})&=\sum_{n=1}^{n_{max}}c_{nl}\phi_{nlm}(\boldsymbol{r}),\\
\phi_{nlm}(\boldsymbol{r})&=N_{nl}r^l e^{-r^2/r_n^2}Y_{lm}(\boldsymbol{\hat{r}}),
\end{align}
$N_{nl}$ is the normalization coefficient with specific radial and orbital angular momentum quantum number $n$, $l$. The size parameter $r_n$ is taken as geometric progression,
\begin{equation}
	r_n=r_1a^{n-1}(n=1,...,n_{max}).
\end{equation}

Through variational principle, the Schr\"odinger equation is transformed to the following matrix equation
\begin{equation}
\sum_{n=1}^{n_{max}} \sum_{n^{\prime}=1}^{n_{max}}(H_{nn^{\prime}}-E_{nl}N_{nn^{\prime}})c_{nl}=0,
\end{equation}
where $H_{nn^{\prime}}$, $N_{nn^{\prime}}$ are matrix elements,
\begin{align}
H_{nn^{\prime}}&=\langle\phi_{n^{\prime}lm}(\boldsymbol{r})|H|\phi_{nlm}(\boldsymbol{r})\rangle,\\
N_{nn^{\prime}}&=\langle\phi_{n^{\prime}lm}(\boldsymbol{r})|\phi_{nlm}(\boldsymbol{r}\rangle.
\end{align}
The mass of tetraquark is then obtained through solving the above generalized eigenvalue problem. Similar to the fixing process in Refs.~\cite{Lu:2020cns,Liu:2019zuc,Liu:2021rtn,Zhang:2022qtp}, the variational parameters $\{r_1,r_n,n_{max}\}$ are adjusted to $\{0.1~\mathrm{fm},5~\mathrm{fm},25\}$ to reach the convergence.
\section{Numerical Results}\label{results}
\subsection{Model Parameters}
The parameters in Cornell potential to be determined are $\alpha_s$, $b$, $\sigma$. We take $b$, $\sigma$ as those in Ref.~\cite{Debastiani:2017msn}, which is fitted by reproducing the mass spectrum of charmonium. The strong coupling constant is computed by
\begin{equation}
\alpha_s\left(Q^2\right)=\frac{4 \pi}{(11-\frac{2}{3}n_f) \ln \frac{Q^2}{\Lambda_{QCD}^2}},
\end{equation}
where the active flavor number $n_f$ is set with $4$, and the QCD scale $\Lambda_{QCD}$ is chosen as $200$ MeV.

There are large uncertainties for the constituent $[cc]$ diquark mass $m_d$ in different models. In potential models~\cite{Debastiani:2017msn,Ferretti:2019zyh,Bedolla:2019zwg}, the charmed $[cc]$ diquark is treated as a composite of two charm quarks bound by the Cornell potential or the GI potential~\cite{Debastiani:2017msn,Bedolla:2019zwg}. Accordingly, the mass of diquark $[cc]$ is calculated with $3130$ MeV and $3329$ MeV in terms of these two potentials. In constituent quark models~\cite{Karliner:2016zzc,Giron:2020wpx,Wu:2022gie}, the fitted $[cc]$ diquark mass varies from $3126.4-3146.4$ MeV to $3300.8$ MeV through the established hidden charmed tetraquark candidates or double charmed baryons. Within QCD sum rule method, the masses of light, heavy-light, and double heavy diquark with different quantum numbers have been computed~\cite{Zhang:2006xp,Kleiv:2013dta,Esau:2019hqw}, and the axial charmed diquark $[cc]$ is predicted with $3.51$ GeV~\cite{Esau:2019hqw}. So far, the diquark mass has not been fixed for lack of experimental data, the constituent $[cc]$ diquark mass is chosen from $3.13$ GeV to $3.51$ GeV in our calculation.

For the parameters in the one boson exchange potential, we take the same values as those in Ref.~\cite{Liu:2019stu} except $f_\omega$, which is proportional to the constituent meson mass in hadronic molecular system.  $f_\omega$ equals $11.7$ when the constituent meson mass is $1.87$ GeV. All the parameters are given in Table~\ref{tab:paras}.
\begin{table}[h]
\caption{\label{tab:paras}
Parameters chosen in this work.}
\begin{ruledtabular}
\begin{tabular}{cccccc}
& Cornell\cite{Debastiani:2017msn} & &OBE\cite{Liu:2019stu}\\
\hline
$b(\mathrm{GeV}^{2})$ & $0.1463$   & $m_\sigma$(MeV) & $600$ \\
$\sigma$(GeV)         & $1.0831$   & $g_\sigma$ & $3.4$\\
 & & $m_{\omega}$(MeV) &$782.6$\\
 & & $\Lambda$(GeV) & $1.0$\\
 & & $g_\omega$ & $2.6$\\
 & & $f_\omega$ & $\frac{11.7}{1.87} m_d$\\
\end{tabular}
\end{ruledtabular}
\end{table}

\subsection{Mass Spectrum in $\mathbf{\bar 3}_c\otimes \mathbf{3}_c$ Color Configuration}
The mass spectrum of $[cc][\bar c\bar c]$ tetraquarks has been calculated with the $[cc]$ diquark mass between $3.13$ GeV and $3.51$ GeV, and the numerical results are given in Table~\ref{tab:3massd}. All possible tetraquarks from $1S$ to $2P$ allowed by symmetry are given in the first column, where the symbols of the tetraquarks in Refs.~\cite{Chao:1980dv,Giron:2020wpx,Debastiani:2017msn} are employed. There are $3$ $S$-wave and $7$ $P$-wave tetraquarks with different internal excitations and $J^{PC}$ quantum numbers. In the table, $n$ and $L$ refer to the radial and orbital angular momentum, $S$ and $J$ refer to the total spin and angular momentum. The $J^{PC}$ quantum numbers are listed in the second column, where $P=(-1)^L$ and $C=(-1)^{L+S}$. The mass spectrum computed with the pure Cornell potential is presented in the third column, while the Cornell potential plus the residual $\sigma$ exchange and $\omega$ exchange potential are presented in the fourth and fifth columns. The mass spectrum from four bodies picture~\cite{Liu:2021rtn} is listed in the last column.

\begin{table*}[htb]

\caption{Mass spectrum (in MeV) of fully-charm tetraquark $[cc][\bar c\bar c]$ in $\mathbf{\bar 3}_c\otimes \mathbf{3}_c$ configuration, where the mass of diquark is set at $3.13-3.51$ GeV.}
\label{tab:3massd}
\begin{ruledtabular}
\begin{tabular}{cccccc}
$n^{2S+1}L_J$ & $J^{PC}$ & Cornell  & $\sigma$ & $\omega$        & Ref.~\cite{Liu:2021rtn} \\
\hline
$1^1S_{0}$    & $0^{++}$ & $6247.9\sim7005.8$         & $6195.1\sim6952.1$         & $6207.0\sim6963.6$         & $6455$                                          \\
$1^3S_{1}$    & $1^{+-}$ & $6275.2\sim7027.6$        & $6223.2\sim6974.6$         & $6251.3\sim7002.7$         & $6500$                                          \\
$1^5S_{2}$    & $2^{++}$ & $6325.4\sim7068.3$         & $6275.2\sim7016.7$         & $6331.1\sim7074.7$         & $6524$                                         \\
$1^1P_{1}$    & $1^{--}$ & $6679.1\sim7419.2$        & $6647.2\sim7385.8$         & $6664.3\sim7403.2$         & $6904$                                          \\
$1^3P_{0}$    & $0^{-+}$ & $6637.3\sim7412.2$         & $6602.2\sim7379.1$         & $6623.0\sim7399.2$         & $6891$                                          \\
$1^3P_{1}$    & $1^{-+}$ & $6680.5\sim7420.3$        & $6648.5\sim7386.9$         & $6671.9\sim7411.1$         & $6908$                                          \\
$1^3P_{2}$    & $2^{-+}$ & $6693.0\sim7431.2$        & $6662.4\sim7399.1$         & $6683.6\sim7421.1$          & $6928$                                          \\
$1^5P_{1}$    & $1^{--}$ & $6624.0\sim7389.3$        & $6587.3\sim7353.7$         & $6623.0\sim7387.9$         & $6768$                                          \\
$1^5P_{2}$    & $2^{--}$ & $6692.6\sim7430.7$        & $6661.7\sim7398.3$         & $6694.0\sim7432.4$         & $6955$                                         \\
$1^5P_{3}$    & $3^{--}$ & $6706.6\sim7443.3$        & $6677.3\sim7412.4$         & $6705.6\sim7442.5$          & $6801$                                          \\
$2^1S_{0}$    & $0^{++}$ & $6777.4\sim7515.2$        & $6752.1\sim7488.7$         & $6763.8\sim7500.9$         & $7185$                                          \\
$2^3S_{1}$    & $1^{+-}$ & $6785.6\sim7521.7$         & $6760.4\sim7495.2$         & $6776.9\sim7512.5$         & $7021$                                          \\
$2^5S_{2}$    & $2^{++}$ & $6802.6\sim7535.0$        & $6777.6\sim7508.7$          & $6803.8\sim7536.3$         & $7032$                                          \\
$2^1P_{1}$    & $1^{--}$ & $7016.7\sim7745.4$        & $6998.0\sim7725.8$         & $7008.0\sim7736.2$         &                                               \\
$2^3P_{0}$    & $0^{-+}$ & $6987.7\sim7748.0$        & $6968.2\sim7729.2$         & $6980.1\sim7740.9$         &                                               \\
$2^3P_{1}$    & $1^{-+}$ & $7017.5\sim7746.1$        & $6998.7\sim7726.3$         & $7012.4\sim7740.7$         &                                               \\
$2^3P_{2}$    & $2^{-+}$ & $7029.0\sim7756.0$        & $7011.0\sim7737.0$         & $7023.5\sim7750.1$         &                                               \\
$2^5P_{1}$    & $1^{--}$ & $6974.1\sim7725.7$         & $6953.6\sim7705.4$         & $6973.6\sim7724.9$         &                                               \\
$2^5P_{2}$    & $2^{--}$ & $7028.4\sim7755.2$        & $7010.2\sim7735.9$         & $7029.4\sim7756.3$         &                                               \\
$2^5P_{3}$    & $3^{--}$ & $7041.3\sim7766.9$        & $7024.0\sim7748.4$         & $7040.9\sim7766.5$          &

\end{tabular}
\end{ruledtabular}
\end{table*}

From Eq.~[\ref{eq:potential}], the mass of $[cc][\bar c\bar c]$ tetraquarks depends explicitly on the constituent $[cc]$ diquark mass. With the additional $\sigma$ boson exchange interaction involved, the mass spectrum of $[cc][\bar c\bar c]$ tetraquarks decreases $40-50$ MeV for the ground $1S$-wave states and $20-30$ MeV for the $2S$-wave and $P$-wave excitations. Obviously, the boson exchange interaction as a residual interaction contributes to the mass spectrum small. The results imply that the additional boson exchange interaction primarily contributes to the attraction of diquark and antidiquark at short distance. With the additional $\omega$ boson exchange interaction involved, the mass spectrum of $[cc][\bar c\bar c]$ tetraquarks decreases negligibly except for the ground $S$-wave states, which implies that the $\omega$ boson exchange interaction play role in much shorter distance. It is similar to the case in hadronic molecular systems~\cite{Liu:2019stu}.

From Eq.~[\ref{eq:potential}], the mass splitting between different states depends on the kinetic energy and the potential. To have a look at how the mass splittings behave, the feature of the splittings is explored. For this reason, the mass splitting between and within different multiplets are calculated and presented in Table~\ref{tab:ls} and Table~\ref{tab:ms}, respectively. In these two tables, the numerical results from the second to the fourth columns are calculated with the pure Cornell potential, additional $\sigma$ boson exchange potential and additional $\omega$ boson exchange potential. As a comparison, the splittings in charmonium system~\cite{ParticleDataGroup:2022pth} are also presented in the last column, which are computed with the central value of the experimental results. In Table~\ref{tab:ls}, a spin-weighted average mass
\begin{equation}
\overline{M_{nL}}=\frac{\sum_J{(2J+1)M_{nL_{J}}}}{\sum_J{(2J+1)}}
\end{equation}
is taken use of. To give the splitting of charmonium in Table~\ref{tab:ls} and Table~\ref{tab:ms}, the data from PDG~\cite{ParticleDataGroup:2022pth} is shown in Table~\ref{tab:mcha}.

\begin{table*}[htb]
\caption{Mass splitting (in MeV) of radial and orbital $[cc][\bar c\bar c]$ with the diquark mass $3.13-3.51$ GeV in different models.}
\label{tab:ls}
\begin{ruledtabular}
\begin{tabular}{ccccccccccc}
 & Cornell & $\sigma$ & $\omega$  & Charmonium~\cite{ParticleDataGroup:2022pth}    \\ \hline
$1P-1S$  & $383.8\sim378.6$  & $403.4\sim398.3$            & $388.1\sim382.8$                                  & $456.7$                                                        \\
$2P-2S$  & $227.7\sim224.3$  & $234.4\sim231.5$            & $228.6\sim225.4$                                 & $\cdots$                                                        \\
$2S-1S$  & $494.1\sim480.6$  & $520.1\sim506.5$            & $499.7\sim486.0$                                           & $605.4$                                                                  \\
$2P-1P$  & $338.0\sim326.4$  & $351.1\sim339.7$            & $340.2\sim328.6$                                                             & $\cdots$

\end{tabular}
\end{ruledtabular}
\end{table*}

\begin{table*}[htb]
\caption{Mass splitting (in MeV) within the same multiplet of $[cc][\bar c\bar c]$ in different models. The minimum and maximum values with the diquark mass $3.13-3.51$ GeV are shown.}
\label{tab:ms}
\begin{ruledtabular}
\begin{tabular}{ccccc}
   & Cornell      & $\sigma$     & $\omega$      & Charmonium~\cite{ParticleDataGroup:2022pth} \\ \hline
$1S$ & $(27.3-77.5)\sim(21.8-62.5)$  & $(28.1-80.1)\sim(22.5-64.6)$  & $(44.3-124.1)\sim(39.1-111.1)$  & $113.0$ \\
$1P$ & $(0.4-82.6)\sim(0.5-54.0)$  & $(0.7-90.0)\sim(0.8-58.7)$  & $(0.0-82.6)\sim(4.0-54.6)$  & $(15.0-141.5)$ \\
$2S$ & $(8.2-25.2)\sim(6.5-19.8)$  & $(8.3-25.5)\sim(6.5-20.0)$  & $(13.1-39.9)\sim(11.6-35.4)$  & $48.4$ \\
$2P$ & $(0.8-67.2)\sim(0.7-41.2)$  & $(0.7-70.4)\sim(0.5-43.0)$  & $(4.4-67.3)\sim(0.2-41.6)$  & $\cdots$
\end{tabular}
\end{ruledtabular}
\end{table*}

\begin{table}[htb]
\caption{Mass spectrum (in MeV) of charmonium from PDG~\cite{ParticleDataGroup:2022pth}.}
\label{tab:mcha}
\begin{ruledtabular}
\begin{tabular}{cccc}
$n^{2S+1}L_J$ & $J^{PC}$ & Name & Mass~\cite{ParticleDataGroup:2022pth} \\ \hline
$1^1S_0$      & $0^{-+}$ & $\eta_c(1S)$     & $2983.9\pm0.4$                                                      \\
$1^3S_1$      & $1^{--}$ &  $J/\psi(1S)$    & $3096.900\pm0.006$                                                      \\
$1^1P_1$      & $1^{+-}$ &  $h_c(1P)$       & $3525.37\pm0.14$                                                     \\
$1^3P_0$      & $0^{++}$ &  $\chi_{c0}(1P)$    & $3414.71\pm0.30$                                                     \\
$1^3P_1$      & $1^{++}$ &  $\chi_{c1}(1P)$    & $3510.67\pm0.05$                                                     \\
$1^3P_2$      & $2^{++}$ &  $\chi_{c2}(1P)$    & $3556.17\pm0.07$                                                     \\
$2^1S_0$      & $0^{-+}$ &  $\eta_c(2S)$    & $3637.7\pm1.1$                                                      \\
$2^3S_1$      & $1^{--}$ &  $\psi(2S)$    & $3686.10\pm0.06$
\end{tabular}
\end{ruledtabular}
\end{table}

From Table~\ref{tab:ls}, the mass splittings of $[cc][\bar c\bar c]$ tetraquarks are $230-520$ MeV for the $1P-1S$, $2P-2S$ and $2S-1S$. The mass splittings for the $1P-1S$ and $2S-1S$ are $70$ MeV and $105$ MeV smaller than those in charmonium. Sine the $2P$ charmonium multiplet has not been definitely established in experiment, relevant splittings in Table~\ref{tab:ls} and Table~\ref{tab:ms} have not estimated. With the boson exchange interactions involved, the mass splittings between different $[cc][\bar c\bar c]$ multiplets increases $7-26$ MeV for the $\sigma$ interaction, and only several MeV for the $\omega$ interaction. The mass splittings of $[cc][\bar c\bar c]$ tetraquarks with the additional $\sigma$ potential are larger than those with the additional $\omega$ potential. In addition, the orbital splitting is smaller than the radial splitting. The mass splittings depend very weakly on the $[cc]$ diquark mass. In fact, the orbital splittings vary couples MeV and the radial splittings vary $\sim10$ MeV when the $[cc]$ diquark mass varies from $3.13$ GeV to $3.51$ GeV. The mass splitting between multiplets is a good experimental observable with small uncertainty.

From Table~\ref{tab:ms}, the largest mass splitting of $[cc][\bar c\bar c]$ within the $1S$ multiplet is about $80$ MeV, and largest mass splitting
within the $2S$ multiplet is about $26$ MeV with additional $\sigma$ potential. The largest mass splitting within the $1S$ multiplet is about $124$ MeV, and largest mass splitting within the $2S$ multiplet is about $40$ MeV with additional $\omega$ potential. The mass splittings within the same multiplet are smaller than those in charmonium system. With the boson exchange interactions involved, the mass splittings within the same $[cc][\bar c\bar c]$ multiplet increase couples MeV. In comparison with the $\sigma$ potential, the $\omega$ potential has a larger effect on the mass splittings within the ground $1S$ state.

The mass splittings within the same multiplet depend also weakly on the $[cc]$ diquark mass, the maximum splitting decreases about $10-30$ MeV when the $[cc]$ diquark mass varies from $3.13$ GeV to $3.51$ GeV. In the constituent quark potential models, the mass splitting within the same multiplet is related to the spin-dependent terms inversely proportional to the constituent mass. So the mass splitting within the same multiplet is smaller in the system with a larger diquark mass.

\section{summary}\label{summary}
In this work, the mass spectrum of fully charmed $[cc][\bar c\bar c]$ tetraquarks from $1S$ to $2P$ excitations has been calculated  in the non-relativistic quark potential model in the diquark-antidiquark picture. In the calculation, the interaction inside the $[cc][\bar c\bar c]$ tetraquarks is modeled as a dominant color interaction between a diquark and an antidiquark, and a residual interaction responsible for the diquark/antidiquark cluster effect. The color interaction between the diquark and the antidiquark is characterized by the conventional Cornell potential, while the residual interaction is modeled as the Yukawa-type scalar $\sigma$ and vector $\omega$ boson exchange potentials.

The contribution of boson exchange interaction to the mass spectrum of $[cc][\bar c\bar c]$ tetraquarks is small. With the additional $\sigma$ boson exchange interaction involved, the mass spectrum of $[cc][\bar c\bar c]$ tetraquarks decreases $40-50$ MeV for the $1S$-wave states, and $20-30$ MeV for the $2S$-wave and $P$-wave excitations. With the additional $\omega$ boson exchange interaction, the mass spectrum of $[cc][\bar c \bar c]$ tetraquarks changes little except for the $S-$wave states. The results imply that the boson exchange interactions are only involved in short distance.

The mass spectrum of $[cc][\bar c \bar c]$ tetraquarks depends on the constituent $[cc]$ diquark mass. Once the $[cc]$ diquark mass be fixed, the mass spectrum of $[cc][\bar c \bar c]$ tetraquarks from $1S$ to $2P$ excitations could be obtained with small uncertainty. Unfortunately, the constituent $[cc]$ diquark mass has not been fixed for certain. In particular, there are large uncertainties for the masses of $[cc][\bar c \bar c]$ candidates in the experiments. We can not assign the observed $X(6600)$, $X(6900)$ and $X(7300)$ based on the absolute numerical results from theory.

For each $S-$wave and $P-$wave excitation, there are three and seven $[cc][\bar c\bar c]$ tetraquarks, respectively, which is different from those in charmonium. Different mass splittings of $[cc][\bar c\bar c]$ tetraquarks have been calculated, and the mass splittings depend weakly on the constituent $[cc]$ diquark mass. With the boson exchange interactions involved, the mass splittings between different $[cc][\bar c\bar c]$ multiplets increases $7-26$ MeV for the $\sigma$ interaction and only several MeV for the $\omega$ interaction, while the mass splittings within the same $[cc][\bar c\bar c]$ multiplet increase only couples MeV. From the mass splittings of $[cc][\bar c \bar c]$ among different excitations, $X(6600)$, $X(6900)$ and $X(7300)$ can not lie in the same radial excitations, and can only lie in different radial excitations. In experiment, to fix the $J^{PC}$ quantum numbers and to measure the mass splittings of $[cc][\bar c\bar c]$ candidates will help people to identify the observed candidates, and will help people to understand of the structure and dynamics of $[cc][\bar c\bar c]$ tetraquarks.

\begin{acknowledgments}
This work is supported by National Natural Science Foundation of China under the grant No. 11975146
\end{acknowledgments}


\begin{thebibliography}{81}%
\makeatletter
\providecommand \@ifxundefined [1]{%
 \@ifx{#1\undefined}
}%
\providecommand \@ifnum [1]{%
 \ifnum #1\expandafter \@firstoftwo
 \else \expandafter \@secondoftwo
 \fi
}%
\providecommand \@ifx [1]{%
 \ifx #1\expandafter \@firstoftwo
 \else \expandafter \@secondoftwo
 \fi
}%
\providecommand \natexlab [1]{#1}%
\providecommand \enquote  [1]{``#1''}%
\providecommand \bibnamefont  [1]{#1}%
\providecommand \bibfnamefont [1]{#1}%
\providecommand \citenamefont [1]{#1}%
\providecommand \href@noop [0]{\@secondoftwo}%
\providecommand \href [0]{\begingroup \@sanitize@url \@href}%
\providecommand \@href[1]{\@@startlink{#1}\@@href}%
\providecommand \@@href[1]{\endgroup#1\@@endlink}%
\providecommand \@sanitize@url [0]{\catcode `\\12\catcode `\$12\catcode `\&12\catcode `\#12\catcode `\^12\catcode `\_12\catcode `\%12\relax}%
\providecommand \@@startlink[1]{}%
\providecommand \@@endlink[0]{}%
\providecommand \url  [0]{\begingroup\@sanitize@url \@url }%
\providecommand \@url [1]{\endgroup\@href {#1}{\urlprefix }}%
\providecommand \urlprefix  [0]{URL }%
\providecommand \Eprint [0]{\href }%
\providecommand \doibase [0]{https://doi.org/}%
\providecommand \selectlanguage [0]{\@gobble}%
\providecommand \bibinfo  [0]{\@secondoftwo}%
\providecommand \bibfield  [0]{\@secondoftwo}%
\providecommand \translation [1]{[#1]}%
\providecommand \BibitemOpen [0]{}%
\providecommand \bibitemStop [0]{}%
\providecommand \bibitemNoStop [0]{.\EOS\space}%
\providecommand \EOS [0]{\spacefactor3000\relax}%
\providecommand \BibitemShut  [1]{\csname bibitem#1\endcsname}%
\let\auto@bib@innerbib\@empty
\bibitem [{\citenamefont {Gell-Mann}(1964)}]{GellMann1964}%
  \BibitemOpen
  \bibfield  {author} {\bibinfo {author} {\bibfnamefont {M.}~\bibnamefont {Gell-Mann}},\ }\bibfield  {title} {\bibinfo {title} {{A Schematic Model of Baryons and Mesons}},\ }\href {https://doi.org/10.1016/S0031-9163(64)92001-3} {\bibfield  {journal} {\bibinfo  {journal} {Phys. Lett.}\ }\textbf {\bibinfo {volume} {8}},\ \bibinfo {pages} {214} (\bibinfo {year} {1964})}\BibitemShut {NoStop}%
\bibitem [{\citenamefont {Choi}\ \emph {et~al.}(2003)\citenamefont {Choi} \emph {et~al.}}]{Belle:2003nnu}%
  \BibitemOpen
  \bibfield  {author} {\bibinfo {author} {\bibfnamefont {S.~K.}\ \bibnamefont {Choi}} \emph {et~al.} (\bibinfo {collaboration} {Belle}),\ }\bibfield  {title} {\bibinfo {title} {{Observation of a narrow charmonium-like state in exclusive $B^\pm \to K^\pm \pi^+ \pi^- J/\psi$ decays}},\ }\href {https://doi.org/10.1103/PhysRevLett.91.262001} {\bibfield  {journal} {\bibinfo  {journal} {Phys. Rev. Lett.}\ }\textbf {\bibinfo {volume} {91}},\ \bibinfo {pages} {262001} (\bibinfo {year} {2003})},\ \Eprint {https://arxiv.org/abs/hep-ex/0309032} {arXiv:hep-ex/0309032} \BibitemShut {NoStop}%
\bibitem [{\citenamefont {Dubynskiy}\ and\ \citenamefont {Voloshin}(2008)}]{Dubynskiy:2008mq}%
  \BibitemOpen
  \bibfield  {author} {\bibinfo {author} {\bibfnamefont {S.}~\bibnamefont {Dubynskiy}}\ and\ \bibinfo {author} {\bibfnamefont {M.~B.}\ \bibnamefont {Voloshin}},\ }\bibfield  {title} {\bibinfo {title} {{Hadro-Charmonium}},\ }\href {https://doi.org/10.1016/j.physletb.2008.07.086} {\bibfield  {journal} {\bibinfo  {journal} {Phys. Lett. B}\ }\textbf {\bibinfo {volume} {666}},\ \bibinfo {pages} {344} (\bibinfo {year} {2008})},\ \Eprint {https://arxiv.org/abs/0803.2224} {arXiv:0803.2224 [hep-ph]} \BibitemShut {NoStop}%
\bibitem [{\citenamefont {Braaten}(2013)}]{Braaten:2013boa}%
  \BibitemOpen
  \bibfield  {author} {\bibinfo {author} {\bibfnamefont {E.}~\bibnamefont {Braaten}},\ }\bibfield  {title} {\bibinfo {title} {{How the $Z_c$(3900) Reveals the Spectra of Quarkonium Hybrid and Tetraquark Mesons}},\ }\href {https://doi.org/10.1103/PhysRevLett.111.162003} {\bibfield  {journal} {\bibinfo  {journal} {Phys. Rev. Lett.}\ }\textbf {\bibinfo {volume} {111}},\ \bibinfo {pages} {162003} (\bibinfo {year} {2013})},\ \Eprint {https://arxiv.org/abs/1305.6905} {arXiv:1305.6905 [hep-ph]} \BibitemShut {NoStop}%
\bibitem [{\citenamefont {Aaij}\ \emph {et~al.}(2020)\citenamefont {Aaij} \emph {et~al.}}]{LHCb:2020bwg}%
  \BibitemOpen
  \bibfield  {author} {\bibinfo {author} {\bibfnamefont {R.}~\bibnamefont {Aaij}} \emph {et~al.} (\bibinfo {collaboration} {LHCb Collaboration}),\ }\bibfield  {title} {\bibinfo {title} {{Observation of structure in the $J /\psi$ -pair mass spectrum}},\ }\href {https://doi.org/10.1016/j.scib.2020.08.032} {\bibfield  {journal} {\bibinfo  {journal} {Sci. Bull.}\ }\textbf {\bibinfo {volume} {65}},\ \bibinfo {pages} {1983} (\bibinfo {year} {2020})},\ \Eprint {https://arxiv.org/abs/2006.16957} {arXiv:2006.16957 [hep-ex]} \BibitemShut {NoStop}%
\bibitem [{\citenamefont {Hayrapetyan}\ \emph {et~al.}(2024)\citenamefont {Hayrapetyan} \emph {et~al.}}]{CMS:2023owd}%
  \BibitemOpen
  \bibfield  {author} {\bibinfo {author} {\bibfnamefont {A.}~\bibnamefont {Hayrapetyan}} \emph {et~al.} (\bibinfo {collaboration} {CMS}),\ }\bibfield  {title} {\bibinfo {title} {{New Structures in the J/\ensuremath{\psi}J/\ensuremath{\psi} Mass Spectrum in Proton-Proton Collisions at s=13\,\,TeV}},\ }\href {https://doi.org/10.1103/PhysRevLett.132.111901} {\bibfield  {journal} {\bibinfo  {journal} {Phys. Rev. Lett.}\ }\textbf {\bibinfo {volume} {132}},\ \bibinfo {pages} {111901} (\bibinfo {year} {2024})},\ \Eprint {https://arxiv.org/abs/2306.07164} {arXiv:2306.07164 [hep-ex]} \BibitemShut {NoStop}%
\bibitem [{\citenamefont {Aad}\ \emph {et~al.}(2023)\citenamefont {Aad} \emph {et~al.}}]{ATLAS:2023bft}%
  \BibitemOpen
  \bibfield  {author} {\bibinfo {author} {\bibfnamefont {G.}~\bibnamefont {Aad}} \emph {et~al.} (\bibinfo {collaboration} {ATLAS}),\ }\bibfield  {title} {\bibinfo {title} {{Observation of an Excess of Dicharmonium Events in the Four-Muon Final State with the ATLAS Detector}},\ }\href {https://doi.org/10.1103/PhysRevLett.131.151902} {\bibfield  {journal} {\bibinfo  {journal} {Phys. Rev. Lett.}\ }\textbf {\bibinfo {volume} {131}},\ \bibinfo {pages} {151902} (\bibinfo {year} {2023})},\ \Eprint {https://arxiv.org/abs/2304.08962} {arXiv:2304.08962 [hep-ex]} \BibitemShut {NoStop}%
\bibitem [{\citenamefont {Iwasaki}(1975)}]{Iwasaki:1975pv}%
  \BibitemOpen
  \bibfield  {author} {\bibinfo {author} {\bibfnamefont {Y.}~\bibnamefont {Iwasaki}},\ }\bibfield  {title} {\bibinfo {title} {{A Possible Model for New Resonances-Exotics and Hidden Charm}},\ }\href {https://doi.org/10.1143/PTP.54.492} {\bibfield  {journal} {\bibinfo  {journal} {Prog. Theor. Phys.}\ }\textbf {\bibinfo {volume} {54}},\ \bibinfo {pages} {492} (\bibinfo {year} {1975})}\BibitemShut {NoStop}%
\bibitem [{\citenamefont {Chao}(1981)}]{Chao:1980dv}%
  \BibitemOpen
  \bibfield  {author} {\bibinfo {author} {\bibfnamefont {K.-T.}\ \bibnamefont {Chao}},\ }\bibfield  {title} {\bibinfo {title} {{The (cc)-($\bar{cc}$) (Diquark - Anti-Diquark) States in $e^+ e^-$ Annihilation}},\ }\href {https://doi.org/10.1007/BF01431564} {\bibfield  {journal} {\bibinfo  {journal} {Z. Phys. C}\ }\textbf {\bibinfo {volume} {7}},\ \bibinfo {pages} {317} (\bibinfo {year} {1981})}\BibitemShut {NoStop}%
\bibitem [{\citenamefont {Ader}\ \emph {et~al.}(1982)\citenamefont {Ader}, \citenamefont {Richard},\ and\ \citenamefont {Taxil}}]{PhysRevD.25.2370}%
  \BibitemOpen
  \bibfield  {author} {\bibinfo {author} {\bibfnamefont {J.~P.}\ \bibnamefont {Ader}}, \bibinfo {author} {\bibfnamefont {J.~M.}\ \bibnamefont {Richard}},\ and\ \bibinfo {author} {\bibfnamefont {P.}~\bibnamefont {Taxil}},\ }\bibfield  {title} {\bibinfo {title} {Do narrow heavy multiquark states exist?},\ }\href {https://doi.org/10.1103/PhysRevD.25.2370} {\bibfield  {journal} {\bibinfo  {journal} {Phys. Rev. D}\ }\textbf {\bibinfo {volume} {25}},\ \bibinfo {pages} {2370} (\bibinfo {year} {1982})}\BibitemShut {NoStop}%
\bibitem [{\citenamefont {Zouzou}\ \emph {et~al.}(1986)\citenamefont {Zouzou}, \citenamefont {Silvestre-Brac}, \citenamefont {Gignoux},\ and\ \citenamefont {Richard}}]{Zouzou1986}%
  \BibitemOpen
  \bibfield  {author} {\bibinfo {author} {\bibfnamefont {S.}~\bibnamefont {Zouzou}}, \bibinfo {author} {\bibfnamefont {B.}~\bibnamefont {Silvestre-Brac}}, \bibinfo {author} {\bibfnamefont {C.}~\bibnamefont {Gignoux}},\ and\ \bibinfo {author} {\bibfnamefont {J.~M.}\ \bibnamefont {Richard}},\ }\bibfield  {title} {\bibinfo {title} {Four-quark bound states},\ }\href {https://doi.org/10.1007/BF01557611} {\bibfield  {journal} {\bibinfo  {journal} {Z. Phys. C}\ }\textbf {\bibinfo {volume} {30}},\ \bibinfo {pages} {457} (\bibinfo {year} {1986})}\BibitemShut {NoStop}%
\bibitem [{\citenamefont {Zhang}(2021)}]{Zhang:2020xtb}%
  \BibitemOpen
  \bibfield  {author} {\bibinfo {author} {\bibfnamefont {J.-R.}\ \bibnamefont {Zhang}},\ }\bibfield  {title} {\bibinfo {title} {{$0^{+}$ fully-charmed tetraquark states}},\ }\href {https://doi.org/10.1103/PhysRevD.103.014018} {\bibfield  {journal} {\bibinfo  {journal} {Phys. Rev. D}\ }\textbf {\bibinfo {volume} {103}},\ \bibinfo {pages} {014018} (\bibinfo {year} {2021})},\ \Eprint {https://arxiv.org/abs/2010.07719} {arXiv:2010.07719 [hep-ph]} \BibitemShut {NoStop}%
\bibitem [{\citenamefont {Wang}(2020)}]{Wang:2020ols}%
  \BibitemOpen
  \bibfield  {author} {\bibinfo {author} {\bibfnamefont {Z.-G.}\ \bibnamefont {Wang}},\ }\bibfield  {title} {\bibinfo {title} {{Tetraquark candidates in the LHCb's di-$J/\psi$ mass spectrum}},\ }\href {https://doi.org/10.1088/1674-1137/abb080} {\bibfield  {journal} {\bibinfo  {journal} {Chin. Phys. C}\ }\textbf {\bibinfo {volume} {44}},\ \bibinfo {pages} {113106} (\bibinfo {year} {2020})},\ \Eprint {https://arxiv.org/abs/2006.13028} {arXiv:2006.13028 [hep-ph]} \BibitemShut {NoStop}%
\bibitem [{\citenamefont {Li}\ \emph {et~al.}(2021)\citenamefont {Li}, \citenamefont {Chang}, \citenamefont {Wang},\ and\ \citenamefont {Wang}}]{Li:2021ygk}%
  \BibitemOpen
  \bibfield  {author} {\bibinfo {author} {\bibfnamefont {Q.}~\bibnamefont {Li}}, \bibinfo {author} {\bibfnamefont {C.-H.}\ \bibnamefont {Chang}}, \bibinfo {author} {\bibfnamefont {G.-L.}\ \bibnamefont {Wang}},\ and\ \bibinfo {author} {\bibfnamefont {T.}~\bibnamefont {Wang}},\ }\bibfield  {title} {\bibinfo {title} {{Mass spectra and wave functions of ${T}_{QQ\bar{Q}\bar{Q}}$ tetraquarks}},\ }\href {https://doi.org/10.1103/PhysRevD.104.014018} {\bibfield  {journal} {\bibinfo  {journal} {Phys. Rev. D}\ }\textbf {\bibinfo {volume} {104}},\ \bibinfo {pages} {014018} (\bibinfo {year} {2021})},\ \Eprint {https://arxiv.org/abs/2104.12372} {arXiv:2104.12372 [hep-ph]} \BibitemShut {NoStop}%
\bibitem [{\citenamefont {Ke}\ \emph {et~al.}(2021)\citenamefont {Ke}, \citenamefont {Han}, \citenamefont {Liu},\ and\ \citenamefont {Shi}}]{Ke:2021iyh}%
  \BibitemOpen
  \bibfield  {author} {\bibinfo {author} {\bibfnamefont {H.-W.}\ \bibnamefont {Ke}}, \bibinfo {author} {\bibfnamefont {X.}~\bibnamefont {Han}}, \bibinfo {author} {\bibfnamefont {X.-H.}\ \bibnamefont {Liu}},\ and\ \bibinfo {author} {\bibfnamefont {Y.-L.}\ \bibnamefont {Shi}},\ }\bibfield  {title} {\bibinfo {title} {{Tetraquark state $X(6900)$ and the interaction between diquark and antidiquark}},\ }\href {https://doi.org/10.1140/epjc/s10052-021-09229-y} {\bibfield  {journal} {\bibinfo  {journal} {Eur. Phys. J. C}\ }\textbf {\bibinfo {volume} {81}},\ \bibinfo {pages} {427} (\bibinfo {year} {2021})},\ \Eprint {https://arxiv.org/abs/2103.13140} {arXiv:2103.13140 [hep-ph]} \BibitemShut {NoStop}%
\bibitem [{\citenamefont {Zhu}(2021)}]{Zhu:2020xni}%
  \BibitemOpen
  \bibfield  {author} {\bibinfo {author} {\bibfnamefont {R.}~\bibnamefont {Zhu}},\ }\bibfield  {title} {\bibinfo {title} {{Fully-heavy tetraquark spectra and production at hadron colliders}},\ }\href {https://doi.org/10.1016/j.nuclphysb.2021.115393} {\bibfield  {journal} {\bibinfo  {journal} {Nucl. Phys. B}\ }\textbf {\bibinfo {volume} {966}},\ \bibinfo {pages} {115393} (\bibinfo {year} {2021})},\ \Eprint {https://arxiv.org/abs/2010.09082} {arXiv:2010.09082 [hep-ph]} \BibitemShut {NoStop}%
\bibitem [{\citenamefont {Bedolla}\ \emph {et~al.}(2020)\citenamefont {Bedolla}, \citenamefont {Ferretti}, \citenamefont {Roberts},\ and\ \citenamefont {Santopinto}}]{Bedolla:2019zwg}%
  \BibitemOpen
  \bibfield  {author} {\bibinfo {author} {\bibfnamefont {M.~A.}\ \bibnamefont {Bedolla}}, \bibinfo {author} {\bibfnamefont {J.}~\bibnamefont {Ferretti}}, \bibinfo {author} {\bibfnamefont {C.~D.}\ \bibnamefont {Roberts}},\ and\ \bibinfo {author} {\bibfnamefont {E.}~\bibnamefont {Santopinto}},\ }\bibfield  {title} {\bibinfo {title} {{Spectrum of fully-heavy tetraquarks from a diquark+antidiquark perspective}},\ }\href {https://doi.org/10.1140/epjc/s10052-020-08579-3} {\bibfield  {journal} {\bibinfo  {journal} {Eur. Phys. J. C}\ }\textbf {\bibinfo {volume} {80}},\ \bibinfo {pages} {1004} (\bibinfo {year} {2020})},\ \Eprint {https://arxiv.org/abs/1911.00960} {arXiv:1911.00960 [hep-ph]} \BibitemShut {NoStop}%
\bibitem [{\citenamefont {Liu}\ \emph {et~al.}(2021)\citenamefont {Liu}, \citenamefont {Liu}, \citenamefont {Zhong},\ and\ \citenamefont {Zhao}}]{Liu:2021rtn}%
  \BibitemOpen
  \bibfield  {author} {\bibinfo {author} {\bibfnamefont {F.-X.}\ \bibnamefont {Liu}}, \bibinfo {author} {\bibfnamefont {M.-S.}\ \bibnamefont {Liu}}, \bibinfo {author} {\bibfnamefont {X.-H.}\ \bibnamefont {Zhong}},\ and\ \bibinfo {author} {\bibfnamefont {Q.}~\bibnamefont {Zhao}},\ }\bibfield  {title} {\bibinfo {title} {{Higher mass spectra of the fully-charmed and fully-bottom tetraquarks}},\ }\href {https://doi.org/10.1103/PhysRevD.104.116029} {\bibfield  {journal} {\bibinfo  {journal} {Phys. Rev. D}\ }\textbf {\bibinfo {volume} {104}},\ \bibinfo {pages} {116029} (\bibinfo {year} {2021})},\ \Eprint {https://arxiv.org/abs/2110.09052} {arXiv:2110.09052 [hep-ph]} \BibitemShut {NoStop}%
\bibitem [{\citenamefont {Dong}\ \emph {et~al.}(2021)\citenamefont {Dong}, \citenamefont {Baru}, \citenamefont {Guo}, \citenamefont {Hanhart},\ and\ \citenamefont {Nefediev}}]{Dong:2020nwy}%
  \BibitemOpen
  \bibfield  {author} {\bibinfo {author} {\bibfnamefont {X.-K.}\ \bibnamefont {Dong}}, \bibinfo {author} {\bibfnamefont {V.}~\bibnamefont {Baru}}, \bibinfo {author} {\bibfnamefont {F.-K.}\ \bibnamefont {Guo}}, \bibinfo {author} {\bibfnamefont {C.}~\bibnamefont {Hanhart}},\ and\ \bibinfo {author} {\bibfnamefont {A.}~\bibnamefont {Nefediev}},\ }\bibfield  {title} {\bibinfo {title} {{Coupled-Channel Interpretation of the LHCb Double-~$J/\psi$~Spectrum and Hints of a New State Near the~ $J/\psi J/\psi$~~Threshold}},\ }\href {https://doi.org/10.1103/PhysRevLett.127.119901} {\bibfield  {journal} {\bibinfo  {journal} {Phys. Rev. Lett.}\ }\textbf {\bibinfo {volume} {126}},\ \bibinfo {pages} {132001} (\bibinfo {year} {2021})},\ \bibinfo {note} {[Erratum: Phys.Rev.Lett. 127, 119901 (2021)]},\ \Eprint {https://arxiv.org/abs/2009.07795} {arXiv:2009.07795 [hep-ph]} \BibitemShut {NoStop}%
\bibitem [{\citenamefont {Wang}\ \emph {et~al.}(2022)\citenamefont {Wang}, \citenamefont {Meng},\ and\ \citenamefont {Oka}}]{Wang:2022yes}%
  \BibitemOpen
  \bibfield  {author} {\bibinfo {author} {\bibfnamefont {G.-J.}\ \bibnamefont {Wang}}, \bibinfo {author} {\bibfnamefont {Q.}~\bibnamefont {Meng}},\ and\ \bibinfo {author} {\bibfnamefont {M.}~\bibnamefont {Oka}},\ }\bibfield  {title} {\bibinfo {title} {{S-wave fully charmed tetraquark resonant states}},\ }\href {https://doi.org/10.1103/PhysRevD.106.096005} {\bibfield  {journal} {\bibinfo  {journal} {Phys. Rev. D}\ }\textbf {\bibinfo {volume} {106}},\ \bibinfo {pages} {096005} (\bibinfo {year} {2022})},\ \Eprint {https://arxiv.org/abs/2208.07292} {arXiv:2208.07292 [hep-ph]} \BibitemShut {NoStop}%
\bibitem [{\citenamefont {Karliner}\ \emph {et~al.}(2017)\citenamefont {Karliner}, \citenamefont {Nussinov},\ and\ \citenamefont {Rosner}}]{Karliner:2016zzc}%
  \BibitemOpen
  \bibfield  {author} {\bibinfo {author} {\bibfnamefont {M.}~\bibnamefont {Karliner}}, \bibinfo {author} {\bibfnamefont {S.}~\bibnamefont {Nussinov}},\ and\ \bibinfo {author} {\bibfnamefont {J.~L.}\ \bibnamefont {Rosner}},\ }\bibfield  {title} {\bibinfo {title} {{$Q Q \bar Q \bar Q$ states: masses, production, and decays}},\ }\href {https://doi.org/10.1103/PhysRevD.95.034011} {\bibfield  {journal} {\bibinfo  {journal} {Phys. Rev. D}\ }\textbf {\bibinfo {volume} {95}},\ \bibinfo {pages} {034011} (\bibinfo {year} {2017})},\ \Eprint {https://arxiv.org/abs/1611.00348} {arXiv:1611.00348 [hep-ph]} \BibitemShut {NoStop}%
\bibitem [{\citenamefont {Wu}\ \emph {et~al.}(2018)\citenamefont {Wu}, \citenamefont {Liu}, \citenamefont {Chen}, \citenamefont {Liu},\ and\ \citenamefont {Zhu}}]{Wu:2016vtq}%
  \BibitemOpen
  \bibfield  {author} {\bibinfo {author} {\bibfnamefont {J.}~\bibnamefont {Wu}}, \bibinfo {author} {\bibfnamefont {Y.-R.}\ \bibnamefont {Liu}}, \bibinfo {author} {\bibfnamefont {K.}~\bibnamefont {Chen}}, \bibinfo {author} {\bibfnamefont {X.}~\bibnamefont {Liu}},\ and\ \bibinfo {author} {\bibfnamefont {S.-L.}\ \bibnamefont {Zhu}},\ }\bibfield  {title} {\bibinfo {title} {{Heavy-flavored tetraquark states with the $QQ\bar{Q}\bar{Q}$ configuration}},\ }\href {https://doi.org/10.1103/PhysRevD.97.094015} {\bibfield  {journal} {\bibinfo  {journal} {Phys. Rev. D}\ }\textbf {\bibinfo {volume} {97}},\ \bibinfo {pages} {094015} (\bibinfo {year} {2018})},\ \Eprint {https://arxiv.org/abs/1605.01134} {arXiv:1605.01134 [hep-ph]} \BibitemShut {NoStop}%
\bibitem [{\citenamefont {Anwar}\ \emph {et~al.}(2018)\citenamefont {Anwar}, \citenamefont {Ferretti}, \citenamefont {Guo}, \citenamefont {Santopinto},\ and\ \citenamefont {Zou}}]{Anwar:2017toa}%
  \BibitemOpen
  \bibfield  {author} {\bibinfo {author} {\bibfnamefont {M.~N.}\ \bibnamefont {Anwar}}, \bibinfo {author} {\bibfnamefont {J.}~\bibnamefont {Ferretti}}, \bibinfo {author} {\bibfnamefont {F.-K.}\ \bibnamefont {Guo}}, \bibinfo {author} {\bibfnamefont {E.}~\bibnamefont {Santopinto}},\ and\ \bibinfo {author} {\bibfnamefont {B.-S.}\ \bibnamefont {Zou}},\ }\bibfield  {title} {\bibinfo {title} {{Spectroscopy and decays of the fully-heavy tetraquarks}},\ }\href {https://doi.org/10.1140/epjc/s10052-018-6073-9} {\bibfield  {journal} {\bibinfo  {journal} {Eur. Phys. J. C}\ }\textbf {\bibinfo {volume} {78}},\ \bibinfo {pages} {647} (\bibinfo {year} {2018})},\ \Eprint {https://arxiv.org/abs/1710.02540} {arXiv:1710.02540 [hep-ph]} \BibitemShut {NoStop}%
\bibitem [{\citenamefont {Debastiani}\ and\ \citenamefont {Navarra}(2019)}]{Debastiani:2017msn}%
  \BibitemOpen
  \bibfield  {author} {\bibinfo {author} {\bibfnamefont {V.~R.}\ \bibnamefont {Debastiani}}\ and\ \bibinfo {author} {\bibfnamefont {F.~S.}\ \bibnamefont {Navarra}},\ }\bibfield  {title} {\bibinfo {title} {{A non-relativistic model for the $[cc][\bar{c}\bar{c}]$ tetraquark}},\ }\href {https://doi.org/10.1088/1674-1137/43/1/013105} {\bibfield  {journal} {\bibinfo  {journal} {Chin. Phys. C}\ }\textbf {\bibinfo {volume} {43}},\ \bibinfo {pages} {013105} (\bibinfo {year} {2019})},\ \Eprint {https://arxiv.org/abs/1706.07553} {arXiv:1706.07553 [hep-ph]} \BibitemShut {NoStop}%
\bibitem [{\citenamefont {Liu}\ \emph {et~al.}(2019{\natexlab{a}})\citenamefont {Liu}, \citenamefont {L\"u}, \citenamefont {Zhong},\ and\ \citenamefont {Zhao}}]{Liu:2019zuc}%
  \BibitemOpen
  \bibfield  {author} {\bibinfo {author} {\bibfnamefont {M.-S.}\ \bibnamefont {Liu}}, \bibinfo {author} {\bibfnamefont {Q.-F.}\ \bibnamefont {L\"u}}, \bibinfo {author} {\bibfnamefont {X.-H.}\ \bibnamefont {Zhong}},\ and\ \bibinfo {author} {\bibfnamefont {Q.}~\bibnamefont {Zhao}},\ }\bibfield  {title} {\bibinfo {title} {{All-heavy tetraquarks}},\ }\href {https://doi.org/10.1103/PhysRevD.100.016006} {\bibfield  {journal} {\bibinfo  {journal} {Phys. Rev. D}\ }\textbf {\bibinfo {volume} {100}},\ \bibinfo {pages} {016006} (\bibinfo {year} {2019}{\natexlab{a}})},\ \Eprint {https://arxiv.org/abs/1901.02564} {arXiv:1901.02564 [hep-ph]} \BibitemShut {NoStop}%
\bibitem [{\citenamefont {L\"u}\ \emph {et~al.}(2020)\citenamefont {L\"u}, \citenamefont {Chen},\ and\ \citenamefont {Dong}}]{Lu:2020cns}%
  \BibitemOpen
  \bibfield  {author} {\bibinfo {author} {\bibfnamefont {Q.-F.}\ \bibnamefont {L\"u}}, \bibinfo {author} {\bibfnamefont {D.-Y.}\ \bibnamefont {Chen}},\ and\ \bibinfo {author} {\bibfnamefont {Y.-B.}\ \bibnamefont {Dong}},\ }\bibfield  {title} {\bibinfo {title} {{Masses of fully heavy tetraquarks $QQ {\bar{Q}} {\bar{Q}}$ in an extended relativized quark model}},\ }\href {https://doi.org/10.1140/epjc/s10052-020-08454-1} {\bibfield  {journal} {\bibinfo  {journal} {Eur. Phys. J. C}\ }\textbf {\bibinfo {volume} {80}},\ \bibinfo {pages} {871} (\bibinfo {year} {2020})},\ \Eprint {https://arxiv.org/abs/2006.14445} {arXiv:2006.14445 [hep-ph]} \BibitemShut {NoStop}%
\bibitem [{\citenamefont {Jin}\ \emph {et~al.}(2020)\citenamefont {Jin}, \citenamefont {Xue}, \citenamefont {Huang},\ and\ \citenamefont {Ping}}]{Jin:2020jfc}%
  \BibitemOpen
  \bibfield  {author} {\bibinfo {author} {\bibfnamefont {X.}~\bibnamefont {Jin}}, \bibinfo {author} {\bibfnamefont {Y.}~\bibnamefont {Xue}}, \bibinfo {author} {\bibfnamefont {H.}~\bibnamefont {Huang}},\ and\ \bibinfo {author} {\bibfnamefont {J.}~\bibnamefont {Ping}},\ }\bibfield  {title} {\bibinfo {title} {{Full-heavy tetraquarks in constituent quark models}},\ }\href {https://doi.org/10.1140/epjc/s10052-020-08650-z} {\bibfield  {journal} {\bibinfo  {journal} {Eur. Phys. J. C}\ }\textbf {\bibinfo {volume} {80}},\ \bibinfo {pages} {1083} (\bibinfo {year} {2020})},\ \Eprint {https://arxiv.org/abs/2006.13745} {arXiv:2006.13745 [hep-ph]} \BibitemShut {NoStop}%
\bibitem [{\citenamefont {Giron}\ and\ \citenamefont {Lebed}(2020)}]{Giron:2020wpx}%
  \BibitemOpen
  \bibfield  {author} {\bibinfo {author} {\bibfnamefont {J.~F.}\ \bibnamefont {Giron}}\ and\ \bibinfo {author} {\bibfnamefont {R.~F.}\ \bibnamefont {Lebed}},\ }\bibfield  {title} {\bibinfo {title} {{Simple spectrum of $c\bar c c\bar c$ states in the dynamical diquark model}},\ }\href {https://doi.org/10.1103/PhysRevD.102.074003} {\bibfield  {journal} {\bibinfo  {journal} {Phys. Rev. D}\ }\textbf {\bibinfo {volume} {102}},\ \bibinfo {pages} {074003} (\bibinfo {year} {2020})},\ \Eprint {https://arxiv.org/abs/2008.01631} {arXiv:2008.01631 [hep-ph]} \BibitemShut {NoStop}%
\bibitem [{\citenamefont {Zhang}\ \emph {et~al.}(2022)\citenamefont {Zhang}, \citenamefont {Wang}, \citenamefont {Li}, \citenamefont {An}, \citenamefont {Deng},\ and\ \citenamefont {Xie}}]{Zhang:2022qtp}%
  \BibitemOpen
  \bibfield  {author} {\bibinfo {author} {\bibfnamefont {J.}~\bibnamefont {Zhang}}, \bibinfo {author} {\bibfnamefont {J.-B.}\ \bibnamefont {Wang}}, \bibinfo {author} {\bibfnamefont {G.}~\bibnamefont {Li}}, \bibinfo {author} {\bibfnamefont {C.-S.}\ \bibnamefont {An}}, \bibinfo {author} {\bibfnamefont {C.-R.}\ \bibnamefont {Deng}},\ and\ \bibinfo {author} {\bibfnamefont {J.-J.}\ \bibnamefont {Xie}},\ }\bibfield  {title} {\bibinfo {title} {{Spectrum of the S-wave fully-heavy tetraquark states}},\ }\href {https://doi.org/10.1140/epjc/s10052-022-11111-4} {\bibfield  {journal} {\bibinfo  {journal} {Eur. Phys. J. C}\ }\textbf {\bibinfo {volume} {82}},\ \bibinfo {pages} {1126} (\bibinfo {year} {2022})},\ \Eprint {https://arxiv.org/abs/2209.13856} {arXiv:2209.13856 [hep-ph]} \BibitemShut {NoStop}%
\bibitem [{\citenamefont {Wang}\ \emph {et~al.}(2023)\citenamefont {Wang}, \citenamefont {Oka},\ and\ \citenamefont {Jido}}]{Wang:2023jqs}%
  \BibitemOpen
  \bibfield  {author} {\bibinfo {author} {\bibfnamefont {G.-J.}\ \bibnamefont {Wang}}, \bibinfo {author} {\bibfnamefont {M.}~\bibnamefont {Oka}},\ and\ \bibinfo {author} {\bibfnamefont {D.}~\bibnamefont {Jido}},\ }\bibfield  {title} {\bibinfo {title} {{Quark confinement for multiquark systems: Application to fully charmed tetraquarks}},\ }\href {https://doi.org/10.1103/PhysRevD.108.L071501} {\bibfield  {journal} {\bibinfo  {journal} {Phys. Rev. D}\ }\textbf {\bibinfo {volume} {108}},\ \bibinfo {pages} {L071501} (\bibinfo {year} {2023})}\BibitemShut {NoStop}%
\bibitem [{\citenamefont {Faustov}\ \emph {et~al.}(2020)\citenamefont {Faustov}, \citenamefont {Galkin},\ and\ \citenamefont {Savchenko}}]{Faustov:2020qfm}%
  \BibitemOpen
  \bibfield  {author} {\bibinfo {author} {\bibfnamefont {R.~N.}\ \bibnamefont {Faustov}}, \bibinfo {author} {\bibfnamefont {V.~O.}\ \bibnamefont {Galkin}},\ and\ \bibinfo {author} {\bibfnamefont {E.~M.}\ \bibnamefont {Savchenko}},\ }\bibfield  {title} {\bibinfo {title} {{Masses of the $QQ\bar Q\bar Q$ tetraquarks in the relativistic diquark--antidiquark picture}},\ }\href {https://doi.org/10.1103/PhysRevD.102.114030} {\bibfield  {journal} {\bibinfo  {journal} {Phys. Rev. D}\ }\textbf {\bibinfo {volume} {102}},\ \bibinfo {pages} {114030} (\bibinfo {year} {2020})},\ \Eprint {https://arxiv.org/abs/2009.13237} {arXiv:2009.13237 [hep-ph]} \BibitemShut {NoStop}%
\bibitem [{\citenamefont {Ida}\ and\ \citenamefont {Kobayashi}(1966)}]{Ida:1966ev}%
  \BibitemOpen
  \bibfield  {author} {\bibinfo {author} {\bibfnamefont {M.}~\bibnamefont {Ida}}\ and\ \bibinfo {author} {\bibfnamefont {R.}~\bibnamefont {Kobayashi}},\ }\bibfield  {title} {\bibinfo {title} {{Baryon resonances in a quark model}},\ }\href {https://doi.org/10.1143/PTP.36.846} {\bibfield  {journal} {\bibinfo  {journal} {Prog. Theor. Phys.}\ }\textbf {\bibinfo {volume} {36}},\ \bibinfo {pages} {846} (\bibinfo {year} {1966})}\BibitemShut {NoStop}%
\bibitem [{\citenamefont {Lichtenberg}\ and\ \citenamefont {Tassie}(1967)}]{Lichtenberg:1967zz}%
  \BibitemOpen
  \bibfield  {author} {\bibinfo {author} {\bibfnamefont {D.~B.}\ \bibnamefont {Lichtenberg}}\ and\ \bibinfo {author} {\bibfnamefont {L.~J.}\ \bibnamefont {Tassie}},\ }\bibfield  {title} {\bibinfo {title} {{Baryon Mass Splitting in a Boson-Fermion Model}},\ }\href {https://doi.org/10.1103/PhysRev.155.1601} {\bibfield  {journal} {\bibinfo  {journal} {Phys. Rev.}\ }\textbf {\bibinfo {volume} {155}},\ \bibinfo {pages} {1601} (\bibinfo {year} {1967})}\BibitemShut {NoStop}%
\bibitem [{\citenamefont {Jaffe}\ and\ \citenamefont {Wilczek}(2003)}]{Jaffe:2003sg}%
  \BibitemOpen
  \bibfield  {author} {\bibinfo {author} {\bibfnamefont {R.~L.}\ \bibnamefont {Jaffe}}\ and\ \bibinfo {author} {\bibfnamefont {F.}~\bibnamefont {Wilczek}},\ }\bibfield  {title} {\bibinfo {title} {{Diquarks and exotic spectroscopy}},\ }\href {https://doi.org/10.1103/PhysRevLett.91.232003} {\bibfield  {journal} {\bibinfo  {journal} {Phys. Rev. Lett.}\ }\textbf {\bibinfo {volume} {91}},\ \bibinfo {pages} {232003} (\bibinfo {year} {2003})},\ \Eprint {https://arxiv.org/abs/hep-ph/0307341} {arXiv:hep-ph/0307341} \BibitemShut {NoStop}%
\bibitem [{\citenamefont {Karliner}\ and\ \citenamefont {Lipkin}(2003)}]{Karliner:2003dt}%
  \BibitemOpen
  \bibfield  {author} {\bibinfo {author} {\bibfnamefont {M.}~\bibnamefont {Karliner}}\ and\ \bibinfo {author} {\bibfnamefont {H.~J.}\ \bibnamefont {Lipkin}},\ }\bibfield  {title} {\bibinfo {title} {{A Diquark - triquark model for the $KN$ pentaquark}},\ }\href {https://doi.org/10.1016/j.physletb.2003.09.062} {\bibfield  {journal} {\bibinfo  {journal} {Phys. Lett. B}\ }\textbf {\bibinfo {volume} {575}},\ \bibinfo {pages} {249} (\bibinfo {year} {2003})},\ \Eprint {https://arxiv.org/abs/hep-ph/0402260} {arXiv:hep-ph/0402260} \BibitemShut {NoStop}%
\bibitem [{\citenamefont {Maiani}\ \emph {et~al.}(2004)\citenamefont {Maiani}, \citenamefont {Piccinini}, \citenamefont {Polosa},\ and\ \citenamefont {Riquer}}]{Maiani:2004uc}%
  \BibitemOpen
  \bibfield  {author} {\bibinfo {author} {\bibfnamefont {L.}~\bibnamefont {Maiani}}, \bibinfo {author} {\bibfnamefont {F.}~\bibnamefont {Piccinini}}, \bibinfo {author} {\bibfnamefont {A.~D.}\ \bibnamefont {Polosa}},\ and\ \bibinfo {author} {\bibfnamefont {V.}~\bibnamefont {Riquer}},\ }\bibfield  {title} {\bibinfo {title} {{A New look at scalar mesons}},\ }\href {https://doi.org/10.1103/PhysRevLett.93.212002} {\bibfield  {journal} {\bibinfo  {journal} {Phys. Rev. Lett.}\ }\textbf {\bibinfo {volume} {93}},\ \bibinfo {pages} {212002} (\bibinfo {year} {2004})},\ \Eprint {https://arxiv.org/abs/hep-ph/0407017} {arXiv:hep-ph/0407017} \BibitemShut {NoStop}%
\bibitem [{\citenamefont {Ferretti}\ \emph {et~al.}(2011)\citenamefont {Ferretti}, \citenamefont {Vassallo},\ and\ \citenamefont {Santopinto}}]{Ferretti:2011zz}%
  \BibitemOpen
  \bibfield  {author} {\bibinfo {author} {\bibfnamefont {J.}~\bibnamefont {Ferretti}}, \bibinfo {author} {\bibfnamefont {A.}~\bibnamefont {Vassallo}},\ and\ \bibinfo {author} {\bibfnamefont {E.}~\bibnamefont {Santopinto}},\ }\bibfield  {title} {\bibinfo {title} {{Relativistic quark-diquark model of baryons}},\ }\href {https://doi.org/10.1103/PhysRevC.83.065204} {\bibfield  {journal} {\bibinfo  {journal} {Phys. Rev. C}\ }\textbf {\bibinfo {volume} {83}},\ \bibinfo {pages} {065204} (\bibinfo {year} {2011})}\BibitemShut {NoStop}%
\bibitem [{\citenamefont {Maiani}\ \emph {et~al.}(2015)\citenamefont {Maiani}, \citenamefont {Polosa},\ and\ \citenamefont {Riquer}}]{Maiani:2015vwa}%
  \BibitemOpen
  \bibfield  {author} {\bibinfo {author} {\bibfnamefont {L.}~\bibnamefont {Maiani}}, \bibinfo {author} {\bibfnamefont {A.~D.}\ \bibnamefont {Polosa}},\ and\ \bibinfo {author} {\bibfnamefont {V.}~\bibnamefont {Riquer}},\ }\bibfield  {title} {\bibinfo {title} {{The New Pentaquarks in the Diquark Model}},\ }\href {https://doi.org/10.1016/j.physletb.2015.08.008} {\bibfield  {journal} {\bibinfo  {journal} {Phys. Lett. B}\ }\textbf {\bibinfo {volume} {749}},\ \bibinfo {pages} {289} (\bibinfo {year} {2015})},\ \Eprint {https://arxiv.org/abs/1507.04980} {arXiv:1507.04980 [hep-ph]} \BibitemShut {NoStop}%
\bibitem [{\citenamefont {Maiani}\ \emph {et~al.}(2018)\citenamefont {Maiani}, \citenamefont {Polosa},\ and\ \citenamefont {Riquer}}]{Maiani:2017kyi}%
  \BibitemOpen
  \bibfield  {author} {\bibinfo {author} {\bibfnamefont {L.}~\bibnamefont {Maiani}}, \bibinfo {author} {\bibfnamefont {A.~D.}\ \bibnamefont {Polosa}},\ and\ \bibinfo {author} {\bibfnamefont {V.}~\bibnamefont {Riquer}},\ }\bibfield  {title} {\bibinfo {title} {{A Theory of $X$ and $Z$ Multiquark Resonances}},\ }\href {https://doi.org/10.1016/j.physletb.2018.01.039} {\bibfield  {journal} {\bibinfo  {journal} {Phys. Lett. B}\ }\textbf {\bibinfo {volume} {778}},\ \bibinfo {pages} {247} (\bibinfo {year} {2018})},\ \Eprint {https://arxiv.org/abs/1712.05296} {arXiv:1712.05296 [hep-ph]} \BibitemShut {NoStop}%
\bibitem [{\citenamefont {Anselmino}\ \emph {et~al.}(1993)\citenamefont {Anselmino}, \citenamefont {Predazzi}, \citenamefont {Ekelin}, \citenamefont {Fredriksson},\ and\ \citenamefont {Lichtenberg}}]{Anselmino:1992vg}%
  \BibitemOpen
  \bibfield  {author} {\bibinfo {author} {\bibfnamefont {M.}~\bibnamefont {Anselmino}}, \bibinfo {author} {\bibfnamefont {E.}~\bibnamefont {Predazzi}}, \bibinfo {author} {\bibfnamefont {S.}~\bibnamefont {Ekelin}}, \bibinfo {author} {\bibfnamefont {S.}~\bibnamefont {Fredriksson}},\ and\ \bibinfo {author} {\bibfnamefont {D.~B.}\ \bibnamefont {Lichtenberg}},\ }\bibfield  {title} {\bibinfo {title} {{Diquarks}},\ }\href {https://doi.org/10.1103/RevModPhys.65.1199} {\bibfield  {journal} {\bibinfo  {journal} {Rev. Mod. Phys.}\ }\textbf {\bibinfo {volume} {65}},\ \bibinfo {pages} {1199} (\bibinfo {year} {1993})}\BibitemShut {NoStop}%
\bibitem [{\citenamefont {Jaffe}(2005)}]{Jaffe:2004ph}%
  \BibitemOpen
  \bibfield  {author} {\bibinfo {author} {\bibfnamefont {R.~L.}\ \bibnamefont {Jaffe}},\ }\bibfield  {title} {\bibinfo {title} {{Exotica}},\ }\href {https://doi.org/10.1016/j.physrep.2004.11.005} {\bibfield  {journal} {\bibinfo  {journal} {Phys. Rept.}\ }\textbf {\bibinfo {volume} {409}},\ \bibinfo {pages} {1} (\bibinfo {year} {2005})},\ \Eprint {https://arxiv.org/abs/hep-ph/0409065} {arXiv:hep-ph/0409065} \BibitemShut {NoStop}%
\bibitem [{\citenamefont {Chen}\ \emph {et~al.}(2016)\citenamefont {Chen}, \citenamefont {Chen}, \citenamefont {Liu},\ and\ \citenamefont {Zhu}}]{Chen:2016qju}%
  \BibitemOpen
  \bibfield  {author} {\bibinfo {author} {\bibfnamefont {H.-X.}\ \bibnamefont {Chen}}, \bibinfo {author} {\bibfnamefont {W.}~\bibnamefont {Chen}}, \bibinfo {author} {\bibfnamefont {X.}~\bibnamefont {Liu}},\ and\ \bibinfo {author} {\bibfnamefont {S.-L.}\ \bibnamefont {Zhu}},\ }\bibfield  {title} {\bibinfo {title} {{The hidden-charm pentaquark and tetraquark states}},\ }\href {https://doi.org/10.1016/j.physrep.2016.05.004} {\bibfield  {journal} {\bibinfo  {journal} {Phys. Rept.}\ }\textbf {\bibinfo {volume} {639}},\ \bibinfo {pages} {1} (\bibinfo {year} {2016})},\ \Eprint {https://arxiv.org/abs/1601.02092} {arXiv:1601.02092 [hep-ph]} \BibitemShut {NoStop}%
\bibitem [{\citenamefont {Lebed}\ \emph {et~al.}(2017)\citenamefont {Lebed}, \citenamefont {Mitchell},\ and\ \citenamefont {Swanson}}]{Lebed:2016hpi}%
  \BibitemOpen
  \bibfield  {author} {\bibinfo {author} {\bibfnamefont {R.~F.}\ \bibnamefont {Lebed}}, \bibinfo {author} {\bibfnamefont {R.~E.}\ \bibnamefont {Mitchell}},\ and\ \bibinfo {author} {\bibfnamefont {E.~S.}\ \bibnamefont {Swanson}},\ }\bibfield  {title} {\bibinfo {title} {{Heavy-Quark QCD Exotica}},\ }\href {https://doi.org/10.1016/j.ppnp.2016.11.003} {\bibfield  {journal} {\bibinfo  {journal} {Prog. Part. Nucl. Phys.}\ }\textbf {\bibinfo {volume} {93}},\ \bibinfo {pages} {143} (\bibinfo {year} {2017})},\ \Eprint {https://arxiv.org/abs/1610.04528} {arXiv:1610.04528 [hep-ph]} \BibitemShut {NoStop}%
\bibitem [{\citenamefont {Olsen}\ \emph {et~al.}(2018)\citenamefont {Olsen}, \citenamefont {Skwarnicki},\ and\ \citenamefont {Zieminska}}]{Olsen:2017bmm}%
  \BibitemOpen
  \bibfield  {author} {\bibinfo {author} {\bibfnamefont {S.~L.}\ \bibnamefont {Olsen}}, \bibinfo {author} {\bibfnamefont {T.}~\bibnamefont {Skwarnicki}},\ and\ \bibinfo {author} {\bibfnamefont {D.}~\bibnamefont {Zieminska}},\ }\bibfield  {title} {\bibinfo {title} {{Nonstandard heavy mesons and baryons: Experimental evidence}},\ }\href {https://doi.org/10.1103/RevModPhys.90.015003} {\bibfield  {journal} {\bibinfo  {journal} {Rev. Mod. Phys.}\ }\textbf {\bibinfo {volume} {90}},\ \bibinfo {pages} {015003} (\bibinfo {year} {2018})},\ \Eprint {https://arxiv.org/abs/1708.04012} {arXiv:1708.04012 [hep-ph]} \BibitemShut {NoStop}%
\bibitem [{\citenamefont {Barabanov}\ \emph {et~al.}(2021)\citenamefont {Barabanov} \emph {et~al.}}]{Barabanov:2020jvn}%
  \BibitemOpen
  \bibfield  {author} {\bibinfo {author} {\bibfnamefont {M.~Y.}\ \bibnamefont {Barabanov}} \emph {et~al.},\ }\bibfield  {title} {\bibinfo {title} {{Diquark correlations in hadron physics: Origin, impact and evidence}},\ }\href {https://doi.org/10.1016/j.ppnp.2020.103835} {\bibfield  {journal} {\bibinfo  {journal} {Prog. Part. Nucl. Phys.}\ }\textbf {\bibinfo {volume} {116}},\ \bibinfo {pages} {103835} (\bibinfo {year} {2021})},\ \Eprint {https://arxiv.org/abs/2008.07630} {arXiv:2008.07630 [hep-ph]} \BibitemShut {NoStop}%
\bibitem [{\citenamefont {Maiani}\ \emph {et~al.}(2005)\citenamefont {Maiani}, \citenamefont {Piccinini}, \citenamefont {Polosa},\ and\ \citenamefont {Riquer}}]{Maiani:2004vq}%
  \BibitemOpen
  \bibfield  {author} {\bibinfo {author} {\bibfnamefont {L.}~\bibnamefont {Maiani}}, \bibinfo {author} {\bibfnamefont {F.}~\bibnamefont {Piccinini}}, \bibinfo {author} {\bibfnamefont {A.~D.}\ \bibnamefont {Polosa}},\ and\ \bibinfo {author} {\bibfnamefont {V.}~\bibnamefont {Riquer}},\ }\bibfield  {title} {\bibinfo {title} {{Diquark-antidiquarks with hidden or open charm and the nature of $X(3872)$}},\ }\href {https://doi.org/10.1103/PhysRevD.71.014028} {\bibfield  {journal} {\bibinfo  {journal} {Phys. Rev. D}\ }\textbf {\bibinfo {volume} {71}},\ \bibinfo {pages} {014028} (\bibinfo {year} {2005})},\ \Eprint {https://arxiv.org/abs/hep-ph/0412098} {arXiv:hep-ph/0412098} \BibitemShut {NoStop}%
\bibitem [{\citenamefont {Brodsky}\ \emph {et~al.}(2014)\citenamefont {Brodsky}, \citenamefont {Hwang},\ and\ \citenamefont {Lebed}}]{Brodsky:2014xia}%
  \BibitemOpen
  \bibfield  {author} {\bibinfo {author} {\bibfnamefont {S.~J.}\ \bibnamefont {Brodsky}}, \bibinfo {author} {\bibfnamefont {D.~S.}\ \bibnamefont {Hwang}},\ and\ \bibinfo {author} {\bibfnamefont {R.~F.}\ \bibnamefont {Lebed}},\ }\bibfield  {title} {\bibinfo {title} {{Dynamical Picture for the Formation and Decay of the Exotic $XYZ$ Mesons}},\ }\href {https://doi.org/10.1103/PhysRevLett.113.112001} {\bibfield  {journal} {\bibinfo  {journal} {Phys. Rev. Lett.}\ }\textbf {\bibinfo {volume} {113}},\ \bibinfo {pages} {112001} (\bibinfo {year} {2014})},\ \Eprint {https://arxiv.org/abs/1406.7281} {arXiv:1406.7281 [hep-ph]} \BibitemShut {NoStop}%
\bibitem [{\citenamefont {Tiwari}\ \emph {et~al.}(2023)\citenamefont {Tiwari}, \citenamefont {Rathaud},\ and\ \citenamefont {Rai}}]{Tiwari:2021tmz}%
  \BibitemOpen
  \bibfield  {author} {\bibinfo {author} {\bibfnamefont {R.}~\bibnamefont {Tiwari}}, \bibinfo {author} {\bibfnamefont {D.~P.}\ \bibnamefont {Rathaud}},\ and\ \bibinfo {author} {\bibfnamefont {A.~K.}\ \bibnamefont {Rai}},\ }\bibfield  {title} {\bibinfo {title} {{Spectroscopy of all charm tetraquark states}},\ }\href {https://doi.org/10.1007/s12648-022-02427-8} {\bibfield  {journal} {\bibinfo  {journal} {Indian J. Phys.}\ }\textbf {\bibinfo {volume} {97}},\ \bibinfo {pages} {943} (\bibinfo {year} {2023})},\ \Eprint {https://arxiv.org/abs/2108.04017} {arXiv:2108.04017 [hep-ph]} \BibitemShut {NoStop}%
\bibitem [{\citenamefont {Yukawa}(1935)}]{Yukawa:1935xg}%
  \BibitemOpen
  \bibfield  {author} {\bibinfo {author} {\bibfnamefont {H.}~\bibnamefont {Yukawa}},\ }\bibfield  {title} {\bibinfo {title} {{On the Interaction of Elementary Particles I}},\ }\href {https://doi.org/10.1143/PTPS.1.1} {\bibfield  {journal} {\bibinfo  {journal} {Proc. Phys. Math. Soc. Jap.}\ }\textbf {\bibinfo {volume} {17}},\ \bibinfo {pages} {48} (\bibinfo {year} {1935})}\BibitemShut {NoStop}%
\bibitem [{\citenamefont {Bryan}\ and\ \citenamefont {Scott}(1964)}]{Bryan:1964zzb}%
  \BibitemOpen
  \bibfield  {author} {\bibinfo {author} {\bibfnamefont {R.~A.}\ \bibnamefont {Bryan}}\ and\ \bibinfo {author} {\bibfnamefont {B.~L.}\ \bibnamefont {Scott}},\ }\bibfield  {title} {\bibinfo {title} {{Nucleon-Nucleon Scattering from One-Boson-Exchange Potentials}},\ }\href {https://doi.org/10.1103/PhysRev.135.B434} {\bibfield  {journal} {\bibinfo  {journal} {Phys. Rev.}\ }\textbf {\bibinfo {volume} {135}},\ \bibinfo {pages} {B434} (\bibinfo {year} {1964})}\BibitemShut {NoStop}%
\bibitem [{\citenamefont {Ball}\ \emph {et~al.}(1966)\citenamefont {Ball}, \citenamefont {Scotti},\ and\ \citenamefont {Wong}}]{Ball:1965sa}%
  \BibitemOpen
  \bibfield  {author} {\bibinfo {author} {\bibfnamefont {J.~S.}\ \bibnamefont {Ball}}, \bibinfo {author} {\bibfnamefont {A.}~\bibnamefont {Scotti}},\ and\ \bibinfo {author} {\bibfnamefont {D.~Y.}\ \bibnamefont {Wong}},\ }\bibfield  {title} {\bibinfo {title} {{One-Boson-Exchange Model of $NN$ and $N\bar N$ Interaction}},\ }\href {https://doi.org/10.1103/PhysRev.142.1000} {\bibfield  {journal} {\bibinfo  {journal} {Phys. Rev.}\ }\textbf {\bibinfo {volume} {142}},\ \bibinfo {pages} {1000} (\bibinfo {year} {1966})}\BibitemShut {NoStop}%
\bibitem [{\citenamefont {Brown}\ \emph {et~al.}(1970)\citenamefont {Brown}, \citenamefont {Downs},\ and\ \citenamefont {Iddings}}]{Brown:1970mj}%
  \BibitemOpen
  \bibfield  {author} {\bibinfo {author} {\bibfnamefont {J.~T.}\ \bibnamefont {Brown}}, \bibinfo {author} {\bibfnamefont {B.~W.}\ \bibnamefont {Downs}},\ and\ \bibinfo {author} {\bibfnamefont {C.~K.}\ \bibnamefont {Iddings}},\ }\bibfield  {title} {\bibinfo {title} {{One-boson-exchange potential model for the $\Lambda N$ interaction}},\ }\href {https://doi.org/10.1016/0003-4916(70)90484-7} {\bibfield  {journal} {\bibinfo  {journal} {Annals Phys.}\ }\textbf {\bibinfo {volume} {60}},\ \bibinfo {pages} {148} (\bibinfo {year} {1970})}\BibitemShut {NoStop}%
\bibitem [{\citenamefont {Partovi}\ and\ \citenamefont {Lomon}(1970)}]{Partovi:1969wd}%
  \BibitemOpen
  \bibfield  {author} {\bibinfo {author} {\bibfnamefont {M.~H.}\ \bibnamefont {Partovi}}\ and\ \bibinfo {author} {\bibfnamefont {E.~L.}\ \bibnamefont {Lomon}},\ }\bibfield  {title} {\bibinfo {title} {{Field theoretical nucleon-nucleon potential}},\ }\href {https://doi.org/10.1103/PhysRevD.2.1999} {\bibfield  {journal} {\bibinfo  {journal} {Phys. Rev. D}\ }\textbf {\bibinfo {volume} {2}},\ \bibinfo {pages} {1999} (\bibinfo {year} {1970})}\BibitemShut {NoStop}%
\bibitem [{\citenamefont {Machleidt}\ \emph {et~al.}(1987)\citenamefont {Machleidt}, \citenamefont {Holinde},\ and\ \citenamefont {Elster}}]{Machleidt:1987hj}%
  \BibitemOpen
  \bibfield  {author} {\bibinfo {author} {\bibfnamefont {R.}~\bibnamefont {Machleidt}}, \bibinfo {author} {\bibfnamefont {K.}~\bibnamefont {Holinde}},\ and\ \bibinfo {author} {\bibfnamefont {C.}~\bibnamefont {Elster}},\ }\bibfield  {title} {\bibinfo {title} {{The Bonn Meson Exchange Model for the Nucleon-Nucleon Interaction}},\ }\href {https://doi.org/10.1016/S0370-1573(87)80002-9} {\bibfield  {journal} {\bibinfo  {journal} {Phys. Rept.}\ }\textbf {\bibinfo {volume} {149}},\ \bibinfo {pages} {1} (\bibinfo {year} {1987})}\BibitemShut {NoStop}%
\bibitem [{\citenamefont {Meissner}(1988)}]{Meissner:1987ge}%
  \BibitemOpen
  \bibfield  {author} {\bibinfo {author} {\bibfnamefont {U.~G.}\ \bibnamefont {Meissner}},\ }\bibfield  {title} {\bibinfo {title} {{Low-Energy Hadron Physics from Effective Chiral Lagrangians with Vector Mesons}},\ }\href {https://doi.org/10.1016/0370-1573(88)90090-7} {\bibfield  {journal} {\bibinfo  {journal} {Phys. Rept.}\ }\textbf {\bibinfo {volume} {161}},\ \bibinfo {pages} {213} (\bibinfo {year} {1988})}\BibitemShut {NoStop}%
\bibitem [{\citenamefont {Holzenkamp}\ \emph {et~al.}(1989)\citenamefont {Holzenkamp}, \citenamefont {Holinde},\ and\ \citenamefont {Speth}}]{Holzenkamp:1989tq}%
  \BibitemOpen
  \bibfield  {author} {\bibinfo {author} {\bibfnamefont {B.}~\bibnamefont {Holzenkamp}}, \bibinfo {author} {\bibfnamefont {K.}~\bibnamefont {Holinde}},\ and\ \bibinfo {author} {\bibfnamefont {J.}~\bibnamefont {Speth}},\ }\bibfield  {title} {\bibinfo {title} {{A Meson Exchange Model for the Hyperon-Nucleon Interaction}},\ }\href {https://doi.org/10.1016/0375-9474(89)90223-6} {\bibfield  {journal} {\bibinfo  {journal} {Nucl. Phys. A}\ }\textbf {\bibinfo {volume} {500}},\ \bibinfo {pages} {485} (\bibinfo {year} {1989})}\BibitemShut {NoStop}%
\bibitem [{\citenamefont {Weinberg}(1991)}]{Weinberg:1991um}%
  \BibitemOpen
  \bibfield  {author} {\bibinfo {author} {\bibfnamefont {S.}~\bibnamefont {Weinberg}},\ }\bibfield  {title} {\bibinfo {title} {{Effective chiral Lagrangians for nucleon-pion interactions and nuclear forces}},\ }\href {https://doi.org/10.1016/0550-3213(91)90231-L} {\bibfield  {journal} {\bibinfo  {journal} {Nucl. Phys. B}\ }\textbf {\bibinfo {volume} {363}},\ \bibinfo {pages} {3} (\bibinfo {year} {1991})}\BibitemShut {NoStop}%
\bibitem [{\citenamefont {Manohar}\ and\ \citenamefont {Georgi}(1984)}]{Manohar:1983md}%
  \BibitemOpen
  \bibfield  {author} {\bibinfo {author} {\bibfnamefont {A.}~\bibnamefont {Manohar}}\ and\ \bibinfo {author} {\bibfnamefont {H.}~\bibnamefont {Georgi}},\ }\bibfield  {title} {\bibinfo {title} {{Chiral Quarks and the Nonrelativistic Quark Model}},\ }\href {https://doi.org/10.1016/0550-3213(84)90231-1} {\bibfield  {journal} {\bibinfo  {journal} {Nucl. Phys. B}\ }\textbf {\bibinfo {volume} {234}},\ \bibinfo {pages} {189} (\bibinfo {year} {1984})}\BibitemShut {NoStop}%
\bibitem [{\citenamefont {{T\"ornqvist, Nils A.}}(1994)}]{Tornqvist:1993ng}%
  \BibitemOpen
  \bibfield  {author} {\bibinfo {author} {\bibnamefont {{T\"ornqvist, Nils A.}}},\ }\bibfield  {title} {\bibinfo {title} {{From the deuteron to deusons, an analysis of deuteron - like meson meson bound states}},\ }\href {https://doi.org/10.1007/BF01413192} {\bibfield  {journal} {\bibinfo  {journal} {Z. Phys. C}\ }\textbf {\bibinfo {volume} {61}},\ \bibinfo {pages} {525} (\bibinfo {year} {1994})},\ \Eprint {https://arxiv.org/abs/hep-ph/9310247} {arXiv:hep-ph/9310247} \BibitemShut {NoStop}%
\bibitem [{\citenamefont {Peter}(1997)}]{Peter:1997me}%
  \BibitemOpen
  \bibfield  {author} {\bibinfo {author} {\bibfnamefont {M.}~\bibnamefont {Peter}},\ }\bibfield  {title} {\bibinfo {title} {{The Static potential in QCD: A Full two loop calculation}},\ }\href {https://doi.org/10.1016/S0550-3213(97)00373-8} {\bibfield  {journal} {\bibinfo  {journal} {Nucl. Phys. B}\ }\textbf {\bibinfo {volume} {501}},\ \bibinfo {pages} {471} (\bibinfo {year} {1997})},\ \Eprint {https://arxiv.org/abs/hep-ph/9702245} {arXiv:hep-ph/9702245} \BibitemShut {NoStop}%
\bibitem [{\citenamefont {Voloshin}\ and\ \citenamefont {Okun}(1976)}]{Voloshin:1976ap}%
  \BibitemOpen
  \bibfield  {author} {\bibinfo {author} {\bibfnamefont {M.~B.}\ \bibnamefont {Voloshin}}\ and\ \bibinfo {author} {\bibfnamefont {L.~B.}\ \bibnamefont {Okun}},\ }\bibfield  {title} {\bibinfo {title} {{Hadron Molecules and Charmonium Atom}},\ }\href@noop {} {\bibfield  {journal} {\bibinfo  {journal} {JETP Lett.}\ }\textbf {\bibinfo {volume} {23}},\ \bibinfo {pages} {333} (\bibinfo {year} {1976})}\BibitemShut {NoStop}%
\bibitem [{\citenamefont {{T\"ornqvist, Nils A.}}(1991)}]{Tornqvist:1991ks}%
  \BibitemOpen
  \bibfield  {author} {\bibinfo {author} {\bibnamefont {{T\"ornqvist, Nils A.}}},\ }\bibfield  {title} {\bibinfo {title} {{Possible large deuteron-like meson meson states bound by pions}},\ }\href {https://doi.org/10.1103/PhysRevLett.67.556} {\bibfield  {journal} {\bibinfo  {journal} {Phys. Rev. Lett.}\ }\textbf {\bibinfo {volume} {67}},\ \bibinfo {pages} {556} (\bibinfo {year} {1991})}\BibitemShut {NoStop}%
\bibitem [{\citenamefont {Valcarce}\ \emph {et~al.}(2005)\citenamefont {Valcarce}, \citenamefont {Garcilazo}, \citenamefont {Fernandez},\ and\ \citenamefont {Gonzalez}}]{Valcarce:2005em}%
  \BibitemOpen
  \bibfield  {author} {\bibinfo {author} {\bibfnamefont {A.}~\bibnamefont {Valcarce}}, \bibinfo {author} {\bibfnamefont {H.}~\bibnamefont {Garcilazo}}, \bibinfo {author} {\bibfnamefont {F.}~\bibnamefont {Fernandez}},\ and\ \bibinfo {author} {\bibfnamefont {P.}~\bibnamefont {Gonzalez}},\ }\bibfield  {title} {\bibinfo {title} {{Quark-model study of few-baryon systems}},\ }\href {https://doi.org/10.1088/0034-4885/68/5/R01} {\bibfield  {journal} {\bibinfo  {journal} {Rept. Prog. Phys.}\ }\textbf {\bibinfo {volume} {68}},\ \bibinfo {pages} {965} (\bibinfo {year} {2005})},\ \Eprint {https://arxiv.org/abs/hep-ph/0502173} {arXiv:hep-ph/0502173} \BibitemShut {NoStop}%
\bibitem [{\citenamefont {Ding}\ \emph {et~al.}(2009)\citenamefont {Ding}, \citenamefont {Liu},\ and\ \citenamefont {Yan}}]{Ding:2009vj}%
  \BibitemOpen
  \bibfield  {author} {\bibinfo {author} {\bibfnamefont {G.-J.}\ \bibnamefont {Ding}}, \bibinfo {author} {\bibfnamefont {J.-F.}\ \bibnamefont {Liu}},\ and\ \bibinfo {author} {\bibfnamefont {M.-L.}\ \bibnamefont {Yan}},\ }\bibfield  {title} {\bibinfo {title} {{Dynamics of Hadronic Molecule in One-Boson Exchange Approach and Possible Heavy Flavor Molecules}},\ }\href {https://doi.org/10.1103/PhysRevD.79.054005} {\bibfield  {journal} {\bibinfo  {journal} {Phys. Rev. D}\ }\textbf {\bibinfo {volume} {79}},\ \bibinfo {pages} {054005} (\bibinfo {year} {2009})},\ \Eprint {https://arxiv.org/abs/0901.0426} {arXiv:0901.0426 [hep-ph]} \BibitemShut {NoStop}%
\bibitem [{\citenamefont {Lee}\ \emph {et~al.}(2009)\citenamefont {Lee}, \citenamefont {Faessler}, \citenamefont {Gutsche},\ and\ \citenamefont {Lyubovitskij}}]{Lee:2009hy}%
  \BibitemOpen
  \bibfield  {author} {\bibinfo {author} {\bibfnamefont {I.~W.}\ \bibnamefont {Lee}}, \bibinfo {author} {\bibfnamefont {A.}~\bibnamefont {Faessler}}, \bibinfo {author} {\bibfnamefont {T.}~\bibnamefont {Gutsche}},\ and\ \bibinfo {author} {\bibfnamefont {V.~E.}\ \bibnamefont {Lyubovitskij}},\ }\bibfield  {title} {\bibinfo {title} {{$X(3872)$ as a molecular $DD^*$ state in a potential model}},\ }\href {https://doi.org/10.1103/PhysRevD.80.094005} {\bibfield  {journal} {\bibinfo  {journal} {Phys. Rev. D}\ }\textbf {\bibinfo {volume} {80}},\ \bibinfo {pages} {094005} (\bibinfo {year} {2009})},\ \Eprint {https://arxiv.org/abs/0910.1009} {arXiv:0910.1009 [hep-ph]} \BibitemShut {NoStop}%
\bibitem [{\citenamefont {Sun}\ \emph {et~al.}(2011)\citenamefont {Sun}, \citenamefont {He}, \citenamefont {Liu}, \citenamefont {Luo},\ and\ \citenamefont {Zhu}}]{Sun:2011uh}%
  \BibitemOpen
  \bibfield  {author} {\bibinfo {author} {\bibfnamefont {Z.-F.}\ \bibnamefont {Sun}}, \bibinfo {author} {\bibfnamefont {J.}~\bibnamefont {He}}, \bibinfo {author} {\bibfnamefont {X.}~\bibnamefont {Liu}}, \bibinfo {author} {\bibfnamefont {Z.-G.}\ \bibnamefont {Luo}},\ and\ \bibinfo {author} {\bibfnamefont {S.-L.}\ \bibnamefont {Zhu}},\ }\bibfield  {title} {\bibinfo {title} {{$Z_b(10610)^\pm$ and $Z_b(10650)^\pm$ as the $B^*\bar{B}$ and $B^*\bar{B}^{*}$ molecular states}},\ }\href {https://doi.org/10.1103/PhysRevD.84.054002} {\bibfield  {journal} {\bibinfo  {journal} {Phys. Rev. D}\ }\textbf {\bibinfo {volume} {84}},\ \bibinfo {pages} {054002} (\bibinfo {year} {2011})},\ \Eprint {https://arxiv.org/abs/1106.2968} {arXiv:1106.2968 [hep-ph]} \BibitemShut {NoStop}%
\bibitem [{\citenamefont {Wang}\ and\ \citenamefont {Wang}(2013)}]{Wang:2013kva}%
  \BibitemOpen
  \bibfield  {author} {\bibinfo {author} {\bibfnamefont {P.}~\bibnamefont {Wang}}\ and\ \bibinfo {author} {\bibfnamefont {X.~G.}\ \bibnamefont {Wang}},\ }\bibfield  {title} {\bibinfo {title} {{Study on X(3872) from effective field theory with pion exchange interaction}},\ }\href {https://doi.org/10.1103/PhysRevLett.111.042002} {\bibfield  {journal} {\bibinfo  {journal} {Phys. Rev. Lett.}\ }\textbf {\bibinfo {volume} {111}},\ \bibinfo {pages} {042002} (\bibinfo {year} {2013})},\ \Eprint {https://arxiv.org/abs/1304.0846} {arXiv:1304.0846 [hep-ph]} \BibitemShut {NoStop}%
\bibitem [{\citenamefont {Guo}\ \emph {et~al.}(2018)\citenamefont {Guo}, \citenamefont {Hanhart}, \citenamefont {Mei\ss{}ner}, \citenamefont {Wang}, \citenamefont {Zhao},\ and\ \citenamefont {Zou}}]{Guo:2017jvc}%
  \BibitemOpen
  \bibfield  {author} {\bibinfo {author} {\bibfnamefont {F.-K.}\ \bibnamefont {Guo}}, \bibinfo {author} {\bibfnamefont {C.}~\bibnamefont {Hanhart}}, \bibinfo {author} {\bibfnamefont {U.-G.}\ \bibnamefont {Mei\ss{}ner}}, \bibinfo {author} {\bibfnamefont {Q.}~\bibnamefont {Wang}}, \bibinfo {author} {\bibfnamefont {Q.}~\bibnamefont {Zhao}},\ and\ \bibinfo {author} {\bibfnamefont {B.-S.}\ \bibnamefont {Zou}},\ }\bibfield  {title} {\bibinfo {title} {{Hadronic molecules}},\ }\href {https://doi.org/10.1103/RevModPhys.90.015004} {\bibfield  {journal} {\bibinfo  {journal} {Rev. Mod. Phys.}\ }\textbf {\bibinfo {volume} {90}},\ \bibinfo {pages} {015004} (\bibinfo {year} {2018})},\ \Eprint {https://arxiv.org/abs/1705.00141} {arXiv:1705.00141 [hep-ph]} \BibitemShut {NoStop}%
\bibitem [{\citenamefont {De~Rujula}\ \emph {et~al.}(1975)\citenamefont {De~Rujula}, \citenamefont {Georgi},\ and\ \citenamefont {Glashow}}]{DeRujula:1975qlm}%
  \BibitemOpen
  \bibfield  {author} {\bibinfo {author} {\bibfnamefont {A.}~\bibnamefont {De~Rujula}}, \bibinfo {author} {\bibfnamefont {H.}~\bibnamefont {Georgi}},\ and\ \bibinfo {author} {\bibfnamefont {S.~L.}\ \bibnamefont {Glashow}},\ }\bibfield  {title} {\bibinfo {title} {{Hadron Masses in a Gauge Theory}},\ }\href {https://doi.org/10.1103/PhysRevD.12.147} {\bibfield  {journal} {\bibinfo  {journal} {Phys. Rev. D}\ }\textbf {\bibinfo {volume} {12}},\ \bibinfo {pages} {147} (\bibinfo {year} {1975})}\BibitemShut {NoStop}%
\bibitem [{\citenamefont {Eichten}\ \emph {et~al.}(1978)\citenamefont {Eichten}, \citenamefont {Gottfried}, \citenamefont {Kinoshita}, \citenamefont {Lane},\ and\ \citenamefont {Yan}}]{Eichten:1978tg}%
  \BibitemOpen
  \bibfield  {author} {\bibinfo {author} {\bibfnamefont {E.}~\bibnamefont {Eichten}}, \bibinfo {author} {\bibfnamefont {K.}~\bibnamefont {Gottfried}}, \bibinfo {author} {\bibfnamefont {T.}~\bibnamefont {Kinoshita}}, \bibinfo {author} {\bibfnamefont {K.~D.}\ \bibnamefont {Lane}},\ and\ \bibinfo {author} {\bibfnamefont {T.-M.}\ \bibnamefont {Yan}},\ }\bibfield  {title} {\bibinfo {title} {{Charmonium: The Model}},\ }\href {https://doi.org/10.1103/PhysRevD.17.3090} {\bibfield  {journal} {\bibinfo  {journal} {Phys. Rev. D}\ }\textbf {\bibinfo {volume} {17}},\ \bibinfo {pages} {3090} (\bibinfo {year} {1978})},\ \bibinfo {note} {[Erratum: Phys.Rev.D 21, 313 (1980)]}\BibitemShut {NoStop}%
\bibitem [{\citenamefont {Godfrey}\ and\ \citenamefont {Isgur}(1985)}]{Godfrey:1985xj}%
  \BibitemOpen
  \bibfield  {author} {\bibinfo {author} {\bibfnamefont {S.}~\bibnamefont {Godfrey}}\ and\ \bibinfo {author} {\bibfnamefont {N.}~\bibnamefont {Isgur}},\ }\bibfield  {title} {\bibinfo {title} {{Mesons in a Relativized Quark Model with Chromodynamics}},\ }\href {https://doi.org/10.1103/PhysRevD.32.189} {\bibfield  {journal} {\bibinfo  {journal} {Phys. Rev. D}\ }\textbf {\bibinfo {volume} {32}},\ \bibinfo {pages} {189} (\bibinfo {year} {1985})}\BibitemShut {NoStop}%
\bibitem [{\citenamefont {Liu}\ \emph {et~al.}(2019{\natexlab{b}})\citenamefont {Liu}, \citenamefont {Wu}, \citenamefont {Pavon~Valderrama}, \citenamefont {Xie},\ and\ \citenamefont {Geng}}]{Liu:2019stu}%
  \BibitemOpen
  \bibfield  {author} {\bibinfo {author} {\bibfnamefont {M.-Z.}\ \bibnamefont {Liu}}, \bibinfo {author} {\bibfnamefont {T.-W.}\ \bibnamefont {Wu}}, \bibinfo {author} {\bibfnamefont {M.}~\bibnamefont {Pavon~Valderrama}}, \bibinfo {author} {\bibfnamefont {J.-J.}\ \bibnamefont {Xie}},\ and\ \bibinfo {author} {\bibfnamefont {L.-S.}\ \bibnamefont {Geng}},\ }\bibfield  {title} {\bibinfo {title} {{Heavy-quark spin and flavor symmetry partners of the $X(3872)$ revisited: What can we learn from the one boson exchange model?}},\ }\href {https://doi.org/10.1103/PhysRevD.99.094018} {\bibfield  {journal} {\bibinfo  {journal} {Phys. Rev. D}\ }\textbf {\bibinfo {volume} {99}},\ \bibinfo {pages} {094018} (\bibinfo {year} {2019}{\natexlab{b}})},\ \Eprint {https://arxiv.org/abs/1902.03044} {arXiv:1902.03044 [hep-ph]} \BibitemShut {NoStop}%
\bibitem [{\citenamefont {Liu}\ \emph {et~al.}(2009)\citenamefont {Liu}, \citenamefont {Luo}, \citenamefont {Liu},\ and\ \citenamefont {Zhu}}]{Liu:2009qhy}%
  \BibitemOpen
  \bibfield  {author} {\bibinfo {author} {\bibfnamefont {X.}~\bibnamefont {Liu}}, \bibinfo {author} {\bibfnamefont {Z.-G.}\ \bibnamefont {Luo}}, \bibinfo {author} {\bibfnamefont {Y.-R.}\ \bibnamefont {Liu}},\ and\ \bibinfo {author} {\bibfnamefont {S.-L.}\ \bibnamefont {Zhu}},\ }\bibfield  {title} {\bibinfo {title} {{$X(3872)$ and Other Possible Heavy Molecular States}},\ }\href {https://doi.org/10.1140/epjc/s10052-009-1020-4} {\bibfield  {journal} {\bibinfo  {journal} {Eur. Phys. J. C}\ }\textbf {\bibinfo {volume} {61}},\ \bibinfo {pages} {411} (\bibinfo {year} {2009})},\ \Eprint {https://arxiv.org/abs/0808.0073} {arXiv:0808.0073 [hep-ph]} \BibitemShut {NoStop}%
\bibitem [{\citenamefont {Park}\ and\ \citenamefont {Lee}(2014)}]{Park:2013fda}%
  \BibitemOpen
  \bibfield  {author} {\bibinfo {author} {\bibfnamefont {W.}~\bibnamefont {Park}}\ and\ \bibinfo {author} {\bibfnamefont {S.~H.}\ \bibnamefont {Lee}},\ }\bibfield  {title} {\bibinfo {title} {{Color spin wave functions of heavy tetraquark states}},\ }\href {https://doi.org/10.1016/j.nuclphysa.2014.02.008} {\bibfield  {journal} {\bibinfo  {journal} {Nucl. Phys. A}\ }\textbf {\bibinfo {volume} {925}},\ \bibinfo {pages} {161} (\bibinfo {year} {2014})},\ \Eprint {https://arxiv.org/abs/1311.5330} {arXiv:1311.5330 [nucl-th]} \BibitemShut {NoStop}%
\bibitem [{\citenamefont {Hiyama}\ \emph {et~al.}(2003)\citenamefont {Hiyama}, \citenamefont {Kino},\ and\ \citenamefont {Kamimura}}]{Hiyama:2003cu}%
  \BibitemOpen
  \bibfield  {author} {\bibinfo {author} {\bibfnamefont {E.}~\bibnamefont {Hiyama}}, \bibinfo {author} {\bibfnamefont {Y.}~\bibnamefont {Kino}},\ and\ \bibinfo {author} {\bibfnamefont {M.}~\bibnamefont {Kamimura}},\ }\bibfield  {title} {\bibinfo {title} {{Gaussian expansion method for few-body systems}},\ }\href {https://doi.org/10.1016/S0146-6410(03)90015-9} {\bibfield  {journal} {\bibinfo  {journal} {Prog. Part. Nucl. Phys.}\ }\textbf {\bibinfo {volume} {51}},\ \bibinfo {pages} {223} (\bibinfo {year} {2003})}\BibitemShut {NoStop}%
\bibitem [{\citenamefont {Ferretti}(2019)}]{Ferretti:2019zyh}%
  \BibitemOpen
  \bibfield  {author} {\bibinfo {author} {\bibfnamefont {J.}~\bibnamefont {Ferretti}},\ }\bibfield  {title} {\bibinfo {title} {{Effective Degrees of Freedom in Baryon and Meson Spectroscopy}},\ }\href {https://doi.org/10.1007/s00601-019-1483-2} {\bibfield  {journal} {\bibinfo  {journal} {Few Body Syst.}\ }\textbf {\bibinfo {volume} {60}},\ \bibinfo {pages} {17} (\bibinfo {year} {2019})}\BibitemShut {NoStop}%
\bibitem [{\citenamefont {Wu}\ and\ \citenamefont {Ma}(2023)}]{Wu:2022gie}%
  \BibitemOpen
  \bibfield  {author} {\bibinfo {author} {\bibfnamefont {T.-W.}\ \bibnamefont {Wu}}\ and\ \bibinfo {author} {\bibfnamefont {Y.-L.}\ \bibnamefont {Ma}},\ }\bibfield  {title} {\bibinfo {title} {{Doubly heavy tetraquark multiplets as heavy antiquark-diquark symmetry partners of heavy baryons}},\ }\href {https://doi.org/10.1103/PhysRevD.107.L071501} {\bibfield  {journal} {\bibinfo  {journal} {Phys. Rev. D}\ }\textbf {\bibinfo {volume} {107}},\ \bibinfo {pages} {L071501} (\bibinfo {year} {2023})},\ \Eprint {https://arxiv.org/abs/2211.15094} {arXiv:2211.15094 [hep-ph]} \BibitemShut {NoStop}%
\bibitem [{\citenamefont {Zhang}\ \emph {et~al.}(2007)\citenamefont {Zhang}, \citenamefont {Huang},\ and\ \citenamefont {Steele}}]{Zhang:2006xp}%
  \BibitemOpen
  \bibfield  {author} {\bibinfo {author} {\bibfnamefont {A.}~\bibnamefont {Zhang}}, \bibinfo {author} {\bibfnamefont {T.}~\bibnamefont {Huang}},\ and\ \bibinfo {author} {\bibfnamefont {T.~G.}\ \bibnamefont {Steele}},\ }\bibfield  {title} {\bibinfo {title} {{Diquark and light four-quark states}},\ }\href {https://doi.org/10.1103/PhysRevD.76.036004} {\bibfield  {journal} {\bibinfo  {journal} {Phys. Rev. D}\ }\textbf {\bibinfo {volume} {76}},\ \bibinfo {pages} {036004} (\bibinfo {year} {2007})},\ \Eprint {https://arxiv.org/abs/hep-ph/0612146} {arXiv:hep-ph/0612146} \BibitemShut {NoStop}%
\bibitem [{\citenamefont {Kleiv}\ \emph {et~al.}(2013)\citenamefont {Kleiv}, \citenamefont {Steele}, \citenamefont {Zhang},\ and\ \citenamefont {Blokland}}]{Kleiv:2013dta}%
  \BibitemOpen
  \bibfield  {author} {\bibinfo {author} {\bibfnamefont {R.~T.}\ \bibnamefont {Kleiv}}, \bibinfo {author} {\bibfnamefont {T.~G.}\ \bibnamefont {Steele}}, \bibinfo {author} {\bibfnamefont {A.}~\bibnamefont {Zhang}},\ and\ \bibinfo {author} {\bibfnamefont {I.}~\bibnamefont {Blokland}},\ }\bibfield  {title} {\bibinfo {title} {{Heavy-light diquark masses from QCD sum rules and constituent diquark models of tetraquarks}},\ }\href {https://doi.org/10.1103/PhysRevD.87.125018} {\bibfield  {journal} {\bibinfo  {journal} {Phys. Rev. D}\ }\textbf {\bibinfo {volume} {87}},\ \bibinfo {pages} {125018} (\bibinfo {year} {2013})},\ \Eprint {https://arxiv.org/abs/1304.7816} {arXiv:1304.7816 [hep-ph]} \BibitemShut {NoStop}%
\bibitem [{\citenamefont {Esau}\ \emph {et~al.}(2019)\citenamefont {Esau}, \citenamefont {Palameta}, \citenamefont {Kleiv}, \citenamefont {Harnett},\ and\ \citenamefont {Steele}}]{Esau:2019hqw}%
  \BibitemOpen
  \bibfield  {author} {\bibinfo {author} {\bibfnamefont {S.}~\bibnamefont {Esau}}, \bibinfo {author} {\bibfnamefont {A.}~\bibnamefont {Palameta}}, \bibinfo {author} {\bibfnamefont {R.~T.}\ \bibnamefont {Kleiv}}, \bibinfo {author} {\bibfnamefont {D.}~\bibnamefont {Harnett}},\ and\ \bibinfo {author} {\bibfnamefont {T.~G.}\ \bibnamefont {Steele}},\ }\bibfield  {title} {\bibinfo {title} {{Axial Vector $cc$ and $bb$ Diquark Masses from QCD Laplace Sum-Rules}},\ }\href {https://doi.org/10.1103/PhysRevD.100.074025} {\bibfield  {journal} {\bibinfo  {journal} {Phys. Rev. D}\ }\textbf {\bibinfo {volume} {100}},\ \bibinfo {pages} {074025} (\bibinfo {year} {2019})},\ \Eprint {https://arxiv.org/abs/1905.12803} {arXiv:1905.12803 [hep-ph]} \BibitemShut {NoStop}%
\bibitem [{\citenamefont {Workman}\ \emph {et~al.}(2022)\citenamefont {Workman} \emph {et~al.}}]{ParticleDataGroup:2022pth}%
  \BibitemOpen
  \bibfield  {author} {\bibinfo {author} {\bibfnamefont {R.~L.}\ \bibnamefont {Workman}} \emph {et~al.} (\bibinfo {collaboration} {Particle Data Group}),\ }\bibfield  {title} {\bibinfo {title} {{Review of Particle Physics}},\ }\href {https://doi.org/10.1093/ptep/ptac097} {\bibfield  {journal} {\bibinfo  {journal} {PTEP}\ }\textbf {\bibinfo {volume} {2022}},\ \bibinfo {pages} {083C01} (\bibinfo {year} {2022})}\BibitemShut {NoStop}%
\end{thebibliography}

%

\end{document}